\documentclass[aps,preprint,showpacs,superscriptaddress,groupedaddress,nofootinbib]{revtex4}  % for double-spaced preprint
\usepackage{graphicx}  % needed for figures
\usepackage{dcolumn}   % needed for some tables
\usepackage{bm,amsmath}        % for math
\usepackage{amssymb}   % for math
\usepackage{color}

\newcommand{\beq}{\begin{equation}} 
\newcommand{\eeq}{\end{equation}} 
\newcommand{\bea}{\begin{eqnarray}} 
\newcommand{\eea}{\end{eqnarray}} 

\begin{document}

%Preprints
\hspace{4.9in} 
\mbox{FERMILAB-PUB-14-053-CD-T}
\vspace{0.2in}
\hspace{6.4in} 
\mbox{IPMU14-0064} \\

\title{Beyond Geolocating: Constraining Higher Dimensional Operators in $H \to 4\ell$ with Off-Shell Production and More}
\author{James~S.~Gainer} \affiliation{Physics Department, University
  of Florida, Gainesville, FL 32611, USA}
\author{Joseph~Lykken} \affiliation{Theoretical Physics Department, Fermilab, Batavia, IL 60510, USA}
\author{Konstantin~T.~Matchev} \affiliation{Physics Department,
  University of Florida, Gainesville, FL 32611, USA}
\author{Stephen Mrenna} \affiliation{SSE Group, Computing Division, Fermilab, Batavia, IL 60510, USA}
\author{Myeonghun~Park} \affiliation{Kavli Institute for the Physics and Mathematics of the Universe (WPI), Todai Institutes for Advanced Study, the University of Tokyo, Japan}
\date{March 19, 2014}

\begin{abstract}
We extend the study of Higgs boson couplings in the ``golden'' 
$gg\to H \to ZZ^\ast \to 4\ell$ channel in two important respects.
First, we demonstrate the importance of off-shell Higgs boson production
($gg\to H^\ast \to ZZ \to 4\ell$)  in determining which operators contribute to the
$HZZ$ vertex.  Second, we include the five operators of lowest non-trivial dimension, including
the $Z_\mu Z^\mu \Box H$ and $H Z_\mu \Box Z^\mu$ operators that are often neglected.
We point out that the former operator can be severely constrained by 
the measurement of the off-shell $H^\ast \to ZZ$ rate and/or 
unitarity considerations. We provide analytic expressions for the off-peak cross-sections 
in the presence of these five operators. On-shell, the $Z_\mu Z^\mu \Box H$
operator is indistinguishable from its Standard Model counterpart $H Z_\mu Z^\mu$,
while the $H Z_\mu \Box Z^\mu$ operator can be probed, in particular, by the $Z^\ast$ 
invariant mass distribution.

\end{abstract}

\maketitle
\section{Introduction}

Now that a Standard Model (SM)-like Higgs boson has
been discovered at the Large Hadron Collider
(LHC)~\cite{oai:arXiv.org:1207.7214, Chatrchyan:2012ufa},
it is critical to measure its couplings.
The sensitivity of the $H \to ZZ^\ast \to 4\ell$
to the couplings of the putative Higgs boson to 
$Z$ bosons is well-established theoretically~\cite{Dell'Aquila:1985ve, Nelson:1986ki, Kniehl:1990yb,
  Soni:1993jc, Chang:1993jy, Barger:1993wt, Arens:1994wd, Choi:2002jk,
  Allanach:2002gn, Buszello:2002uu, Schalla:2004ura, Godbole:2007cn,
  Kovalchuk:2008zz, Keung:2008ve, Antipin:2008hj, Cao:2009ah,
  Gao:2010qx, DeRujula:2010ys, Englert:2010ud, Matsuzaki:2011ch,
  DeSanctis:2011yc, logan, Gainer:2011xz, oai:arXiv.org:1110.4405,
  Englert:2012ct, Campbell:2012cz, Campbell:2012ct, Kauer:2012hd,
  Kniehl:2012rz, Moffat:2012pb, Coleppa:2012eh, Bolognesi:2012mm,
  Boughezal:2012tz, Stolarski:2012ps, Cea:2012ud, Kumar:2012ba,
  Geng:2012hy, Avery:2012um, Masso:2012eq, Chen:2012jy, Modak:2013sb,
  Kanemura:2013mc, Gainer:2013rxa, Isidori:2013cla, Frank:2013gca,
  Grinstein:2013vsa, Caola:2013yja, Banerjee:2013apa, Sun:2013yra,
  Anderson:2013afp, Chen:2013waa, Buchalla:2013mpa, Chen:2013ejz,
  Campbell:2013una, Chen:2014pia, Gonzalez-Alonso:2014rla}; 
measurements of this channel have indeed been
performed by the experimental collaborations~\cite{Chatrchyan:2012jja,
  Aad:2013xqa, Chatrchyan:2013mxa, 1494488, 1523699, 1523767, 1542341,
  1542387}.

Recently, the importance of the off-shell cross section ($M_{4\ell} \gg 125$ GeV)
for measuring the full width of the Higgs boson has been
demonstrated~\cite{Caola:2013yja, Campbell:2013una,
  oai:arXiv.org:1206.4803, Kauer:2013cga, Campbell:2013wga,
  Passarino:2013bha}.
We point out that the off-shell cross section in this channel is
also useful for constraining anomalous $HZZ$ couplings, since
these anomalous operators are of a higher dimension and can enhance the production
cross section at large values of the invariant mass. 
Previous studies of Higgs boson couplings at the LHC (see, e.g.,~\cite{Dawson:2013bba}
and references therein) have focused on three specific operators, one of mass
dimension three and two of mass dimension five.  Here we also study two 
additional dimension five operators that are suppressed on
shell~\cite{Gainer:2013rxa} (see also  Refs.~\cite{Contino:2013kra,
  Artoisenet:2013puc, Boudjema:2013qla, Alloul:2013naa,Contino:2014aaa}).

In Section~\ref{amp and lag}, we 
discuss parametrizations of the $XZZ$ couplings
(we consider an arbitrary scalar, $X$, in our discussions).
Five independent operators
(or equivalently, five independent Lorentz structures in the amplitude)
should be considered. The measurement of the couplings of these five 
operators is the cornerstone of the future LHC physics program 
and will proceed in several stages:
\begin{enumerate}
\item The measurement of the overall signal rate in the four-lepton channel
from an on-shell Higgs boson provides an important constraint on these five operators,
effectively reducing the parameter space by one dimension \cite{Gainer:2013rxa}.
This ``geolocating'' procedure is reviewed and extended to five degrees of freedom in
Section~\ref{geo}.
\item The measurement of the Higgs boson contribution to the 
$ZZ$ continuum at high invariant masses provides a second, independent
constraint that is the subject of Section~\ref{off-shell}.
\item Finally, precision measurements of decay kinematics on the Higgs boson peak 
provide additional information on the tensor structure of the
$XZZ$ couplings, as discussed in Section~\ref{on-shell}.
\end{enumerate}
We present our conclusions in Section~\ref{Conclusions}.

\section{Parametrization of $XZZ$ Couplings}
\label{amp and lag}

%In the study of the $X \to ZZ^\ast \to 4\ell$ channel, 
There are two
obvious and equivalent approaches to describing the coupling of an arbitrary
spin-zero scalar to two $Z$ bosons:
\begin{itemize}
\item introducing a
general amplitude for $X \to Z_{\lambda_1} Z_{\lambda_2}$,\footnote{The
$Z$ bosons have arbitrary invariant masses.  We will not assume any $Z$
boson to be on-shell, unless explicitly noted.} as is done,
e.g., in Refs.~\cite{Gao:2010qx, DeRujula:2010ys, Bolognesi:2012mm},
\item or through the operators in an effective theory
Lagrangian.  
\end{itemize}
The correspondence between these two prescriptions is:
\begin{eqnarray}
\label{lag 1}
i~\epsilon_1^\ast \cdot \epsilon_2^\ast & ~\Longleftrightarrow~ & -\frac{1}{2} X Z_\mu Z^\mu,   \\ 
\label{lag 2}
i~(p_1 \cdot p_2) (\epsilon_1^\ast \cdot \epsilon_2^\ast) &
~\Longleftrightarrow~ & \frac{1}{2} X \partial_\mu
Z_\nu \partial^\mu Z^\nu, \\
\label{lag 3}
i~(p_1 \cdot \epsilon_2^\ast) (p_2 \cdot \epsilon_1^\ast) &
~\Longleftrightarrow~ & \frac{1}{2} X \partial_\mu
Z_\nu \partial^\nu Z^\mu, \\ 
\label{lag 4}
i~\epsilon_{\mu\nu\rho\sigma} \epsilon_1^{\ast,\mu}
\epsilon_2^{\ast,\nu} p_1^\rho p_2^\sigma & ~\Longleftrightarrow~ &
-\frac{1}{2}\epsilon_{\mu\nu\rho\sigma} \partial^\mu
Z^\nu \partial^\rho Z^\sigma, \\ 
\label{lag 5} 
i~(p_1^2 + p_2^2)(\epsilon_1^\ast \cdot \epsilon_2^\ast) &
~\Longleftrightarrow~ & X Z_\mu \Box Z^\mu, 
\end{eqnarray} 
where $\epsilon_1^\mu = \epsilon^\mu(p_1)$ and $\epsilon_2^\mu =
\epsilon^\mu(p_2)$ are gauge boson polarization vectors.
The five operators (\ref{lag 1}-\ref{lag 5}) are
dimension five or less.\footnote{If we assume that the overall constant
contains one power of the vacuum expectation value, we must refer to, e.g., a dimension five
operator as a dimension six operator.}  These operators correspond to
the five independent amplitude structures which have mass dimension
two or less.

In either approach, there is the freedom to
choose the most convenient set of operators as a basis for a particular
application.
Our basis is described below:
\begin{itemize}
\item The expression (\ref{lag 1}) is proportional the tree-level SM Higgs
boson coupling.  For convenience, we therefore define
\begin{equation}
\mathcal{O}_1 = -\frac{M_Z^2}{v} X Z_\mu Z^\mu,
\end{equation}
where $v$ is the Higgs vacuum expectation value, $246$ GeV; hence
$\mathcal{O}_1$ is equal to the tree-level SM coupling. 
\item
Of the five operators, only
(\ref{lag 4}) is invariant under the gauge transformation $Z_\mu
\to Z_\mu + \partial_\mu \chi$.\footnote{Invariance under the full set of
$SU(2) \times U(1)$ gauge transformations depends on the coefficients
of the corresponding operators in $X \to WW$, $X \to Z\gamma$, 
and $X \to \gamma\gamma$. As we are only considering $X \to ZZ$
channels, we will use the term ``gauge invariant'' to
mean invariant under $Z_\mu \to Z_\mu + \partial_\mu \chi$.}
We therefore define 
\begin{equation}
\mathcal{O}_3 = -\frac{1}{2 v} X F_{\mu\nu} \tilde{F}^{\mu\nu}
\label{Percy}
\end{equation}
to be proportional to this expression, where $\tilde{F}_{\mu\nu}
=\frac{1}{2} \epsilon_{\mu \nu \rho \sigma} F^{\rho \sigma}$
and $F_{\mu\nu} = \partial_\mu Z_\nu - \partial_\nu Z_\mu$.  
\item
None of the remaining four operators in (\ref{lag 1}-\ref{lag 5}) are individually gauge invariant,
but the difference of expressions (\ref{lag 2}) and
(\ref{lag 3}) is.  We therefore define $\mathcal{O}_2$ to be
proportional to this difference:
\begin{equation}
\mathcal{O}_2 =-\frac{1}{2 v} X F_{\mu\nu} F^{\mu\nu}.
\end{equation}
\end{itemize}

In Ref.~\cite{Gainer:2013rxa}, we presented a framework for measuring
the couplings of the putative
Higgs boson $X$ to a pair of gauge bosons with
a primary focus on  the ``golden''  $X \to Z Z^\ast \to 4\ell$
channel.  In that work, we considered in detail only $\mathcal{O}_1$,
$\mathcal{O}_2$, and $\mathcal{O}_3$ and described how, after fixing
the overall rate, the measurements of the coefficients of these
operators corresponded to the ``geolocation'' of the Higgs boson couplings
on a suitably defined sphere.
In this work, we will explore the phenomenological
consequences of performing such measurements in the full
five-dimensional operator space, in particular considering operators
which were mentioned, but ultimately neglected, in
Ref.~\cite{Gainer:2013rxa}.  

Before proceeding, we note that, in general, complex contributions to the 
form factors in the amplitude
can be generated through loops involving light particles; schemes for
measuring the coupling in such scenarios were discussed in
Ref.~\cite{Gainer:2013rxa}.
However, such loop-induced contributions are expected to be
small (see, e.g. Ref~\cite{Chen:2013waa}).   All couplings are 
taken to be real in the analysis presented here.

To study the phenomenological consequences of 
the full five-dimensional operator space, we must
first identify the two basis operators 
not space spanned by
$\mathcal{O}_1, \mathcal{O}_2$, and $\mathcal{O}_3$.
A convenient choice, for phenomenological reasons, is:
\begin{equation}
\mathcal{O}_5 = \frac{2}{v} X Z_\mu \square Z^\mu,
\end{equation}
which is proportional to the operator in expression (\ref{lag 5}).
For the final basis operator, one choice is:
\begin{equation}
\mathcal{O}_4 =  \frac{M_Z^2}{M_X^2 v}\Box X Z_\mu Z^\mu,
\end{equation}
where $M_X$ is the mass of the putative Higgs boson ($\approx 125$ GeV).
This operator is equivalent to the operators in expressions (\ref{lag
  2}) and (\ref{lag 5}) after using integration by parts.  Specifically
\begin{eqnarray}
\mathcal{O}_4 & ~\Longleftrightarrow & \frac{M_Z^2}{M_X^2 v} X
( \partial_\mu Z_\nu \partial^\mu Z^\nu + Z_\mu \Box Z^\mu),
\end{eqnarray}
which can be seen directly by considering the corresponding amplitudes.
As an alternative to $\mathcal{O}_4$, we will also consider an
operator which is proportional to the sum of the
operators in expressions (\ref{lag 2}) and (\ref{lag 3}) and hence is
orthogonal to $\mathcal{O}_2$.  We define this operator as
\begin{equation}
\mathcal{O}_6 = \frac{1}{v} X \left(\partial_\mu Z_\nu \partial^\nu
  Z^\mu+ \partial_\mu Z_\nu \partial^\mu Z^\nu\right).
\label{Harvin}  
\end{equation}
Note that $\{\mathcal{O}_1, \mathcal{O}_2, \mathcal{O}_3,
\mathcal{O}_4, \mathcal{O}_5\}$ and $\{\mathcal{O}_1, \mathcal{O}_2, \mathcal{O}_3,
\mathcal{O}_5, \mathcal{O}_6\}$ are bases, but $\{\mathcal{O}_1, \mathcal{O}_2, \mathcal{O}_3,
\mathcal{O}_4, \mathcal{O}_5, \mathcal{O}_6\}$ is a linearly
dependent set.

%%%%%%%%%%%%% Beginning OF TABLE ################%%%%%%%%%%%%%
\begin{table}
\begin{tabular}{| c | c | c | c |}
\hline
Operator & Dimension & CP   &  Gauge invariant \\
\hline\hline
$\mathcal{O}_1$ & 3 & even & No \\ \hline
$\mathcal{O}_2$ & 5 & even & Yes \\ \hline
$\mathcal{O}_3$ & 5 & odd  &  Yes \\ \hline\hline
$\mathcal{O}_4$ & 5 & even & No \\ \hline
$\mathcal{O}_5$ & 5 & even & No \\ \hline
\end{tabular}
\caption{\label{tab:properties} A summary of the properties of the $\mathcal{O}_i$
operators considered in the text.}
\end{table}
%%%%%%%%%%%%% End OF TABLE ################%%%%%%%%%%%%%

Choosing $\{\mathcal{O}_1, \mathcal{O}_2, \mathcal{O}_3,
\mathcal{O}_4, \mathcal{O}_5\}$ as our basis, we obtain the Lagrangian
\begin{eqnarray}
\mathcal{L} \supset
\sum_{i=1}^5 \kappa_i \mathcal{O}_i = 
-\kappa_1 \frac{M_Z^2}{v} X Z_\mu Z^\mu
-\frac{\kappa_2}{2 v} X F_{\mu\nu} F^{\mu\nu}
-\frac{\kappa_3}{2 v} X F_{\mu\nu} \tilde{F}^{\mu\nu} \\
+\frac{\kappa_4 M_Z^2}{M_X^2 v} \Box X Z_\mu Z^\mu
+\frac{2\kappa_5}{v} X Z_\mu \square Z^\mu.
\label{Lag}
\end{eqnarray}
The amplitude corresponding to this Lagrangian may be written as
\beq
\mathcal{A}=-\frac{2i}{v}\epsilon_1^{*\mu} \epsilon_2^{*\nu}\left(a_1
  g_{\mu\nu}+a_2 p_{1\nu} p_{2\mu}
+a_3 \epsilon_{\mu\nu\rho\sigma} \,p_1^\rho p_2^\sigma 
\right) \, ,
\eeq
where
\bea
a_1 &\equiv& \kappa_1 M_Z^2+(2 (M_Z^2/M_X^2)\kappa_4-\kappa_2) p_1
\cdot p_2 +((M_Z^2/M_X^2)\kappa_4 + \kappa_5) (p_1^2+p_2^2) , \\
a_2 &\equiv& \kappa_2, \\
a_3 &\equiv& \kappa_3. 
\label{amp prime} 
\eea
Different operators (or equivalently, different amplitude structures) correspond to 
different symmetry properties, as is elucidated in Table~\ref{tab:properties}.
Thus, for example, the most general $CP$-even coupling involves the four operators 
$\mathcal{O}_1$, $\mathcal{O}_2$, $\mathcal{O}_4$, and $\mathcal{O}_5$.  The
most general gauge-invariant coupling involves only $\mathcal{O}_2$ and $\mathcal{O}_3$.
We emphasize also that this choice of operators allows one to parametrize all 
amplitude structures up to a given mass dimension.  In particular, 
$\kappa_1$, $\kappa_2$, $\kappa_4$, and
$\kappa_5$ can parametrize any Bose symmetric, Lorentz invariant
kinematic function with mass dimension $\le$ 2 for $a_1$, while
retaining sufficient freedom to assign any possible constant value to
$a_2.$\footnote{See Ref.~\cite{Chen:2013waa} for a dictionary of
  conventions used for describing $XZZ$ couplings in various works.}

\section{Geolocating: The On-Peak Cross Section}\label{geo}

In Ref.~\cite{Gainer:2013rxa}, we provided a parameterization of $XZZ$
couplings in terms of directions on a suitably defined sphere with a
constant value for the on-peak ($M_{4\ell} = M_X$)
cross section times branching ratio for the $4\ell$
final state.  We note in passing that this 
``geolocating''
approach has the experimental benefit of making the
normalization of the differential cross section used in the
Matrix Element Method~\cite{kondo1, kondo2, kondo3, 
dalitz, oai:arXiv.org:hep-ex/9808029, abazov, vigil, canelli,
Gainer:2013iya} trivial.
To obtain the analogous ``sphere'' in the five-dimensional 
$\kappa_i$ space corresponding to the Lagrangian in
Eq.~(\ref{Lag}), we must determine the coefficients
$\gamma_{ij}$ in the equation
\begin{equation}
\Gamma(X\to ZZ \to 4\ell) = \Gamma_{SM} \sum_{i,j}\gamma_{ij}\kappa_i\kappa_j,
\label{GammaZZ}
\end{equation}
where $\Gamma(X\to ZZ \to 4\ell)$ is the partial width for $X \to
ZZ^\ast \to 4\ell$ for the given final state ($4e$, $4\mu$ or $2e2\mu$) after specified selections,
$\Gamma_{SM}$ is the value of this quantity for the tree-level SM
($\kappa_i = \delta_{i1}$), and the $\kappa_i$ are
defined by Eq.~(\ref{Lag}).
We take $\gamma_{ij} = \gamma_{ji}$.

%%%%%%%%%%%%% Beginning OF TABLE ################%%%%%%%%%%%%%
\begin{table}
\begin{tabular}{| c | c | c | c | c | c | c | c |}
\hline
$\gamma_{11} = \gamma_{14} = \gamma_{44}$ & $\gamma_{22}$ &
$\gamma_{12} = \gamma_{24}$ & 
$\gamma_{33}$ & $\gamma_{13}$ =
$\gamma_{23}$ = $\gamma_{34}$ = $\gamma_{35}$ & $\gamma_{25}$ &
$\gamma_{15}$ = $\gamma_{45}$ & $\gamma_{55}$ \\
\hline 
$1$ & $0.090$ & $-0.250$ & $0.038$ & $0$ & $-0.250$ & $0.978$ & $0.987$ \\ \hline
\hline
\end{tabular}
\caption{\label{gamma table} Numerical values for the coefficients
defined in Eq.~(\ref{GammaZZ}) that give the partial width for decay
of the putative Higgs boson to the $2e2\mu$ final state with no event selection applied.}
\end{table}
%%%%%%%%%%%%% End OF TABLE ################%%%%%%%%%%%%%

For any kinematic configuration with
$M_{4\ell} = M_X$, the contributions to the amplitude from
$\mathcal{O}_1$ and from $\mathcal{O}_4$ are equal.
Thus
\begin{equation}
\gamma_{1j} = \gamma_{4j},
\end{equation}
and in particular $\gamma_{11} = \gamma_{14} = \gamma_{44} = 1$
(as $\gamma_{11} = 1$ by construction). 
Also, as the interference between parity odd and parity even
amplitudes generically vanishes at the level of total cross sections,
$\gamma_{3j} = 0$ for $j \ne 3$.
Thus, the only $\gamma_{ij}$ which we need to calculate, beyond those
provided in Ref.~\cite{Gainer:2013rxa}, are $\gamma_{15},
\gamma_{25}$, and $\gamma_{55}$.  For convenience, we present all
$\gamma_{ij}$ for the $2e2\mu$ final state without event selection in
Table~\ref{gamma table}.  In general these values depend both on the
choice of four-lepton final state and the event selection applied.

It is interesting that $\gamma_{55}$ is close to, but slightly less
than, $1$.  We therefore explore how this value arises.  
In general,
\begin{equation}
\frac{\Gamma_B}{\Gamma_A} = \frac{1}{\Gamma_A} \int \frac{d
  \Gamma_B}{d\mathbf{x}} d\mathbf{x} = 
\int \bigg( \frac{d \Gamma_B}{d\mathbf{x}}
\bigg/\frac{d\Gamma_A}{d\mathbf{x}}\bigg)
\bigg( \frac{d\Gamma_A}{d\mathbf{x}} \bigg/
\Gamma_A\bigg) d\mathbf{x}
= \left\langle
\bigg( \frac{d \Gamma_B}{d\mathbf{x}}
\bigg/\frac{d\Gamma_A}{d\mathbf{x}}\bigg)
\right \rangle_A,
\end{equation}
that is, the ratio of widths is given by the expectation value of the
ratio of differential widths as found using the appropriate
hypothesis.  If $d\Gamma_i/d\mathbf{x}$ is the differential width for
some set of kinematic variables, $\mathbf{x}$, when
$\kappa_i = 1$ and $\kappa_j = 0$ for $j \ne i$, then we find
\begin{equation}
\bigg( \frac{d \Gamma_5}{d\mathbf{x}}
\bigg/\frac{d\Gamma_1}{d\mathbf{x}}\bigg)
=
\bigg(\frac{M_{Z_1}^2 + M_{Z_2}^2}{M_Z^2}\bigg)^2,
\label{Tim}
\end{equation}
where $M_{Z_{1(2)}}$ is the invariant mass of the heavier (lighter)
lepton pair.  Thus 
\begin{equation}
\gamma_{55} = 
\left\langle
\bigg(\frac{M_{Z_1}^2 + M_{Z_2}^2}{M_Z^2}\bigg)^2
\right \rangle_{SM}.
\label{Tebow}
\end{equation}
As for most events with $M_{4\ell} \approx M_X$, 
$M_{Z_1} \approx M_Z$ and $M_{Z_2} \lesssim M_X -
M_Z$, so with $M_X = 125$ one would expect $(M_{Z_1}^2 + M_{Z_2}^2)^2
/ M_Z^4 \approx 1.1 - 1.3$, which disagrees with our result for
$\gamma_{55}$ in Table~\ref{gamma table}.  
%%%%%%%%%%%%% Beginning OF FIGURE ################%%%%%%%%%%%%%
\begin{figure}[t]
\includegraphics[width=0.75 \textwidth]{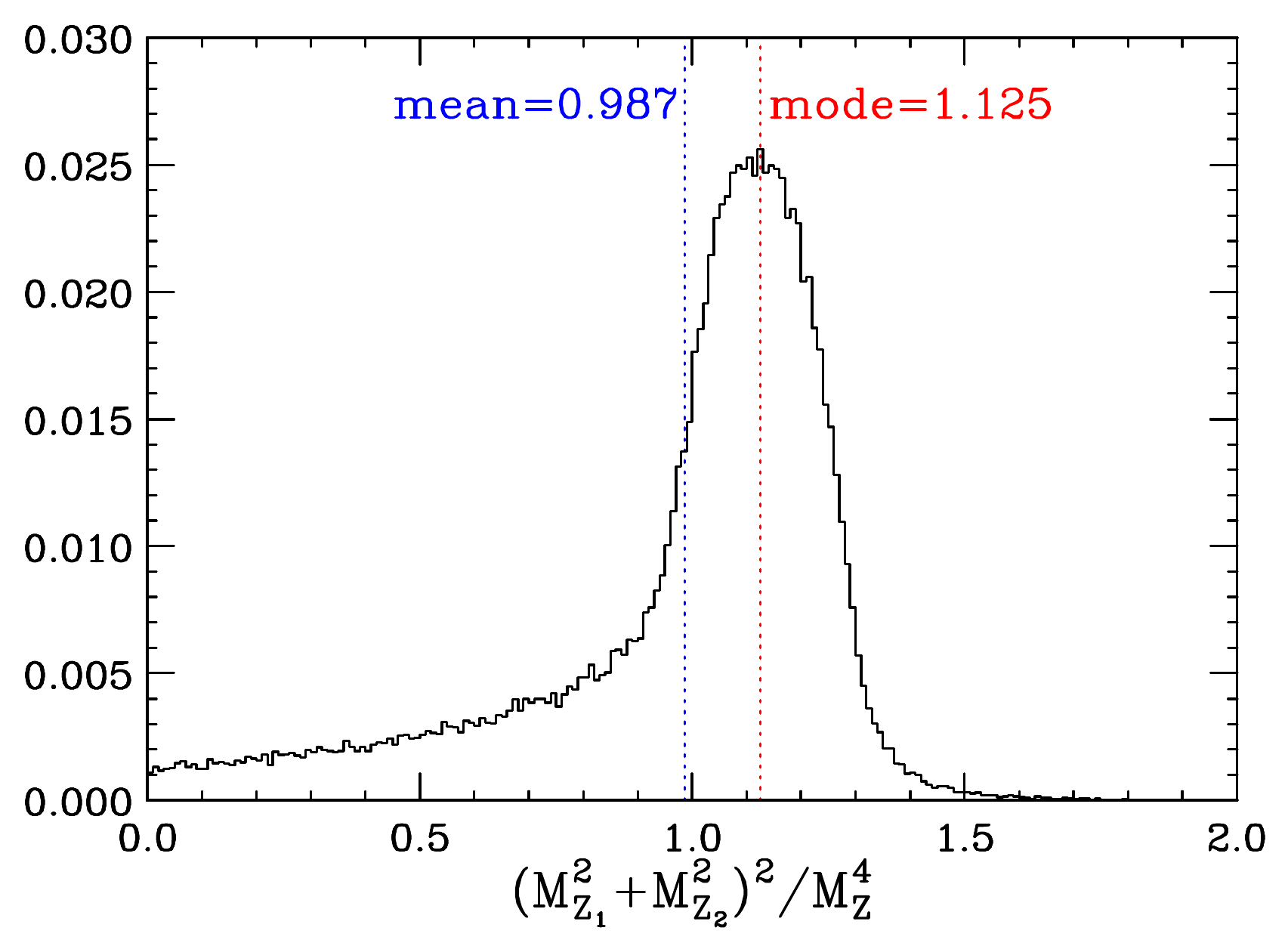} 
%\vspace{-36 pt}
\caption{The distribution of the quantity
  $(M_{Z_1}^2 + M_{Z_2}^2)^2/ M_Z^4$, which is the ratio of
  differential cross sections due to the operator $\mathcal{O}_5$ and
  due to the SM operator, $\mathcal{O}_1$, as evaluated for SM events (see Eqs.~(\ref{Tim}) and (\ref{Tebow})).
  The mean of this quantity is equal to $\gamma_{55}$. \label{fig:gamma_55_hist}}
\end{figure}
%%%%%%%%%%%%% End OF FIGURE %%%%%%%%%%%%%%%%%%%%%%%%%%%%%%%%%
However, this is a naive expectation.   
Fig.~\ref{fig:gamma_55_hist} illustrates 
that while the peak of the distribution of $(M_{Z_1}^2 + M_{Z_2}^2)^2
/ M_Z^4$ for SM events is $1.125$, a long tail extends to
very low values of this quantity.  This tail lowers the average value
of the quantity, and hence of $\gamma_{55}$ to $0.987$, as shown in
Table~\ref{gamma table}.  We note that in this paper we utilize the
event generators MadGraph5~\cite{Alwall:2011uj} and 
CalcHEP~\cite{Belyaev:2012qa} using a model file created with 
FeynRules~\cite{Alloul:2013bka}.

We have presented the $\gamma_{ij}$ corresponding to a particular Higgs boson
width in the limit of no event selection; a more realistic analysis should include the
event selection, 
efficiencies, etc.  We emphasize that the three operators ($\mathcal{O}_1$, 
$\mathcal{O}_2$, and $\mathcal{O}_3$) that were the focus in
Ref.~\cite{Gainer:2013rxa}, and which have been the focus of most experimental
and theoretical analyses thus far do not exhaust all the possibilities.  Even
if studies of these three operators seem to indicate a SM-like Higgs boson, one must 
still probe the complementary $(\kappa_1,\kappa_4,\kappa_5)$ space to conclusively establish the 
boson's identity.

\section{Off-Shell Phenomenology of $XZZ$ Operators}\label{off-shell}

\subsection{Invariant Mass Dependence of Off-Shell Cross Sections}

As noted above, there has been much interest recently in using
four-lepton events from off-shell Higgs boson production, i.e.,~events with
$M_{4\ell} \gg M_X$, 
to constrain the total Higgs boson width~\cite{Caola:2013yja, Campbell:2013una,
  oai:arXiv.org:1206.4803, Kauer:2013cga, Campbell:2013wga,
  Passarino:2013bha}.
We point out here that, for a fixed value of the $X\to ZZ$ partial width (\ref{GammaZZ}) (or
sphere of fixed radius in geolocating language), the off-shell $X^*\to ZZ$
cross section due to any of the dimension five operators (\ref{Percy}-\ref{Harvin})
is much higher than in the Standard Model.
The experimental sensitivity to this off-shell production is
greatly enhanced through interference with the NLO $gg \to ZZ$ 
background~\cite{Caola:2013yja, Campbell:2013una,
  oai:arXiv.org:1206.4803, Kauer:2013cga, Campbell:2013wga,
  Passarino:2013bha, Glover:1988rg, Matsuura:1991pj, Zecher:1994kb,
  Bern:1997sc, Binoth:2005ua, Anastasiou:2007mz, Grazzini:2008tf,
  Binoth:2008pr, Campbell:2011bn},
so determining the precise experimental sensitivity to some
non-standard $XZZ$ couplings is somewhat nontrivial.  In this paper,
we consider only the enhancement in cross
sections relative to the Standard Model that is attained with these
operators; a detailed study of the sensitivity, including the effects
of interference, will be treated in future work.

Before proceeding, we consider the obvious question of
what value of the $ggX$ coupling to use.  In the Standard Model, the
$ggX$ coupling
is given by
\begin{eqnarray}
g_{ggX}(M_{4\ell}) = \frac{\alpha_{s}(M_{4\ell})}
{4\,\pi\,v} \sum_{Q} A_{1/2}^H(\tau_Q),
\label{ggH-evolve}
\end{eqnarray}
at one loop, where
\begin{eqnarray}
A_{1/2}^H(\tau) & = & 2 [\tau +(\tau -1)f(\tau)]\, \tau^{-2},
\label{eq:A}
\end{eqnarray}
$f(\tau)$ is defined by
\begin{eqnarray}
  f(\tau)=\left\{
  \begin{array}{ll}  \displaystyle
  \arcsin^2\sqrt{\tau} & \tau\leq 1 \\
  \displaystyle -\frac{1}{4}\left[ \log\frac{1+\sqrt{1-\tau^{-1}}}
    {1-\sqrt{1-\tau^{-1}}}-i\pi \right]^2 \hspace{0.5cm} & \tau>1
\end{array} \right\},
\label{eq:f(tau)}
\end{eqnarray} 
and $\tau_Q = M_{4\ell}^2/4 M_Q^2$,
following the expressions in e.g. Refs.~\cite{Gunion:1989we, Djouadi:2005gi}.
This expression, more frequently viewed as describing the evolution
of the $ggH$ coupling with $M_H$, can be interpreted somewhat more
generally as it gives the value of this coupling at a particular value
of invariant mass, regardless of the on-shell mass of the resonance.

\begin{figure}[t]
\includegraphics[width=0.95 \textwidth]{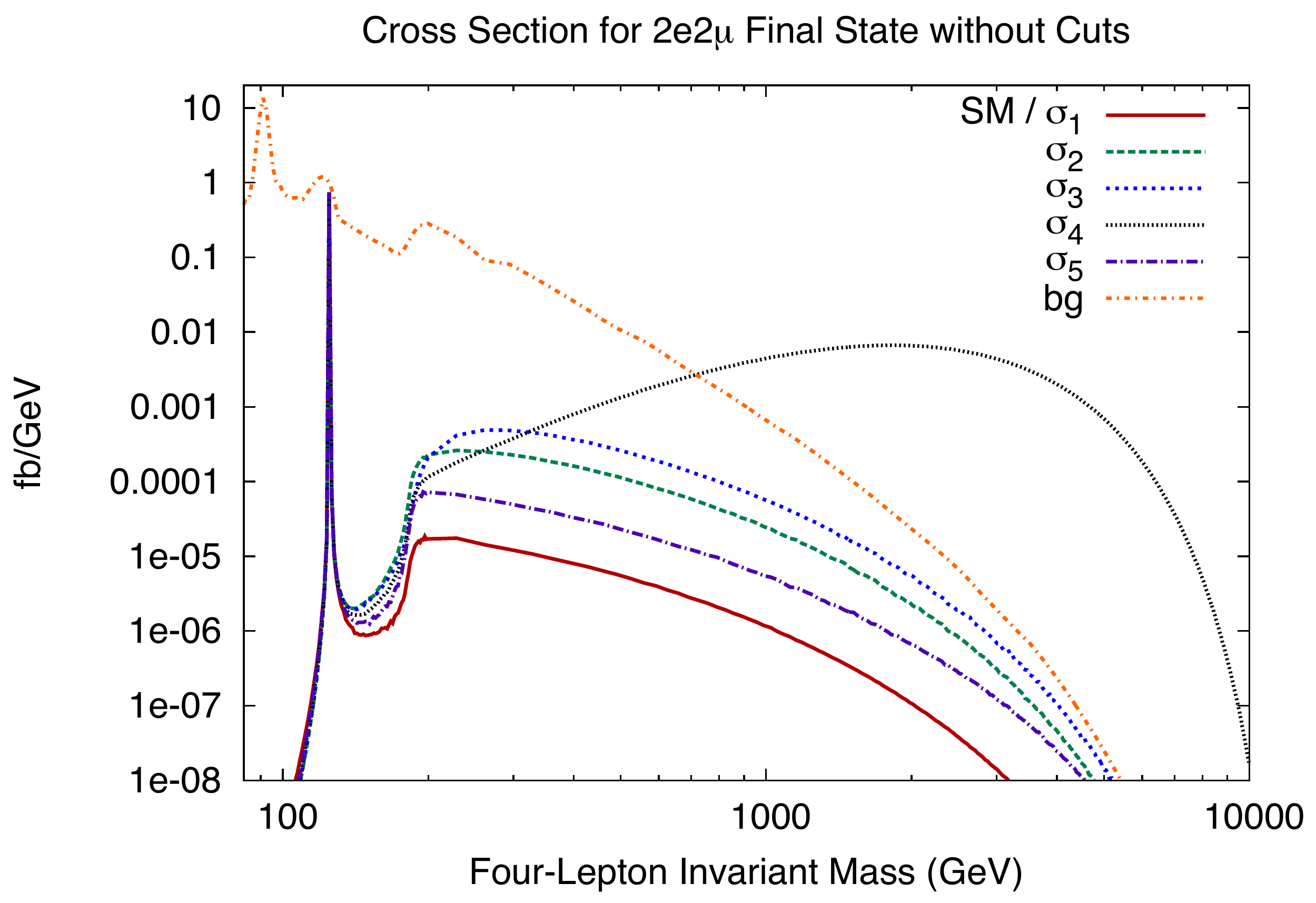} % Here is how to
                                % import EPS art
%\vspace{-36 pt}
\caption{The differential cross section as a function
  of four-lepton invariant mass for $2e2\mu$ events before event selections.
  Results are shown for pure $\mathcal{O}_1$, $\mathcal{O}_2$, $\mathcal{O}_3$,
  $\mathcal{O}_4$, and $\mathcal{O}_5$ 
  couplings (cf. Eq.~(\ref{Lag})), as well as
  for the irreducible $q\bar{q} \to ZZ \to 2e2\mu$ background (bg).  
  There is no event selection applied to the signal events; for the background, a minimal
  $M_{l\bar{l}} > 1$ GeV selection is applied to avoid infrared divergences. 
  For
  each signal hypothesis, the normalization has been chosen to be
  equal to the entire SM on-peak Higgs boson cross section in this channel.
  In this figure, the $ggX$ coupling is taken to be
  constant with respect to invariant mass. \label{fig:2e2mu-xsec-ggX-fixed}}
\end{figure}

\begin{figure}[t]
\includegraphics[width=0.95 \textwidth]{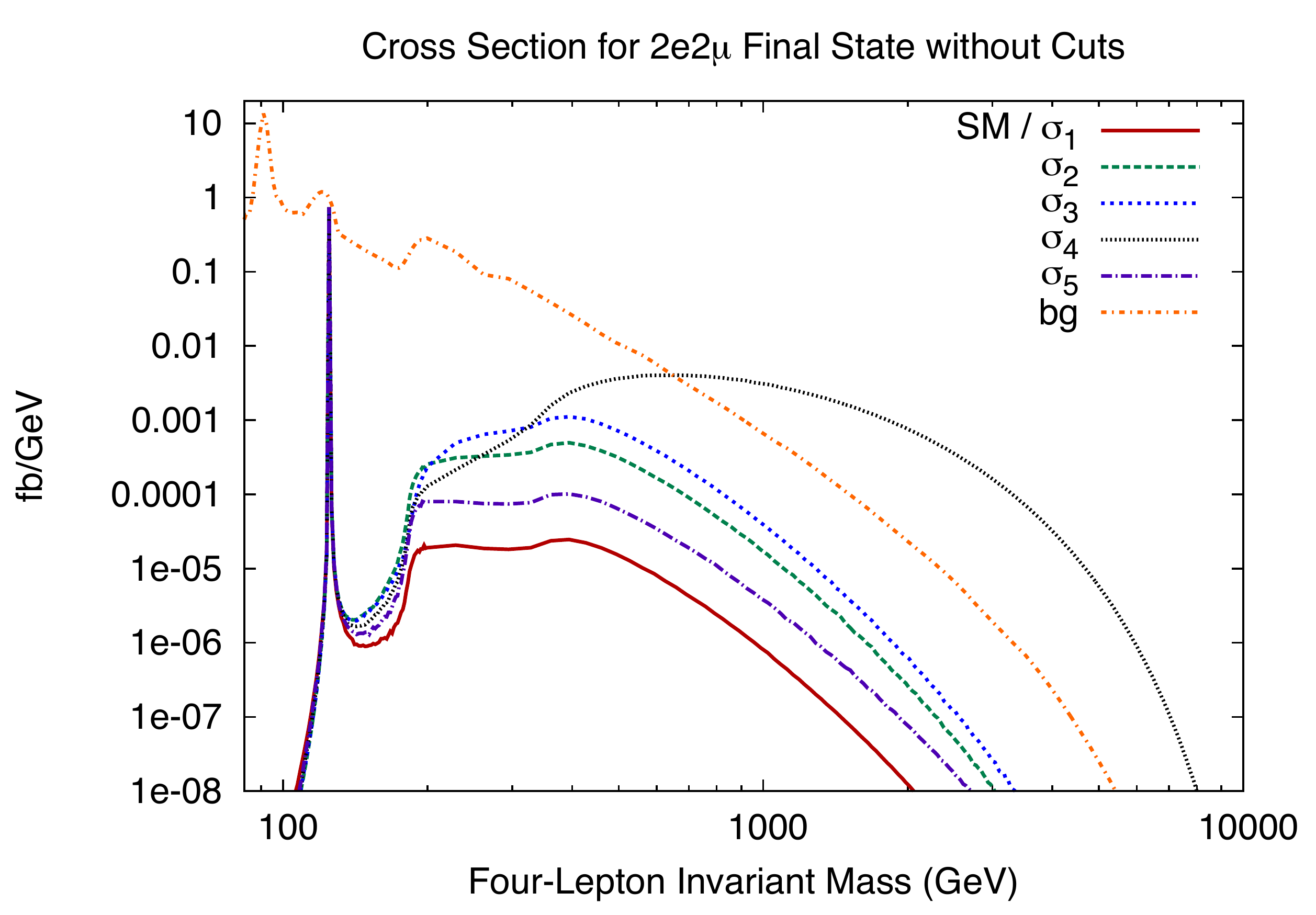} % Here is how to
                                % import EPS art
%\vspace{-36 pt}
\caption{The same as Fig.~\ref{fig:2e2mu-xsec-ggX-fixed}, but in this figure, the $ggX$
  coupling evolves 
  with invariant mass according to the
  expression in Eq.~(\ref{ggH-evolve}).
  \label{fig:2e2mu-xsec-ggX-evolve}}
\end{figure}

However, if we are introducing (in some cases radically) new physics
in the $XZZ$ couplings, we cannot necessarily assume that the SM
expression for the $ggX$ coupling will hold.  Therefore, we consider
an alternative hypothesis that the $ggX$ coupling is fixed at all
scales to its SM
value at $125$ GeV.
We show the LO cross sections $\sigma_{1-5}$ as a function of
$M_{4\ell}$ for the five ``pure'' operators $\mathcal{O}_{1-5}$ in
Fig.~\ref{fig:2e2mu-xsec-ggX-fixed}, in which the $ggX$ coupling does not
evolve with $M_{4\ell}$.  In Fig.~\ref{fig:2e2mu-xsec-ggX-evolve}, we
show these same cross sections, but now calculated with a 
coupling that evolves according to Eq.~(\ref{ggH-evolve}). 
Explicitly, $\sigma_i$ is the
cross section, in a particular $ggX$ coupling scenario, 
when $\kappa_i = \gamma_{ii}^{-1/2}$ and $\kappa_j = 0$ for $i \ne j$.
This choice of $\kappa_i$ serves to normalize the cross sections, so
that the SM value for cross section times branching ratio for $M_{4\ell} \approx
125$ GeV is obtained.
%%%%%%%%%%%%% Beginning OF FIGURE ################%%%%%%%%%%%%%
%%%%%%%%%%%%% End OF FIGURE %%%%%%%%%%%%%%%%%%%%%%%%%%%%%%%%%
%%%%%%%%%%%%% Beginning OF FIGURE ################%%%%%%%%%%%%%
%%%%%%%%%%%%% End OF FIGURE %%%%%%%%%%%%%%%%%%%%%%%%%%%%%%%%%
Signal and background rates integrated over a range of off-shell
invariant masses are provided in Table~\ref{tab:xsec-table}. 
We note from this table, and from Figs.~\ref{fig:2e2mu-xsec-ggX-fixed} 
and~\ref{fig:2e2mu-xsec-ggX-evolve} above,
that $\sigma_{2-5}$ are significantly larger than $\sigma_1$, the
SM off-shell cross section, though the overall scale of cross sections
is relatively small, with the exception of $\sigma_4$.
While, as noted above, we cannot
translate these observations directly into a sensitivity, largely
because
of the importance of interference with the $gg \to ZZ$ continuum
background, it is clear
that the off-shell cross sections provide a source of information
about the tensor $XZZ$ couplings that is complementary to data
obtained on the Higgs boson mass peak.  
As the large values of $\sigma_4$ are symptomatic of potential
unitarity-violating behavior, in Subsection~\ref{The U} we will quantify the reduction of the cross section for
$\mathcal{O}_4$ when one only integrates over values of invariant mass
consistent with unitarity requirements. 
%%%%%%%%%%%%% Beginning OF TABLE ################%%%%%%%%%%%%%
\begin{table}
\begin{tabular}{| l || c | c | c | c |}
\hline
Operator & $\sigma > M_X$, fixed $g_{ggX}$ 
& $\sigma > 250$ GeV, fixed
$g_{ggX}$
& $\sigma > M_X$, $g_{ggX}(M_{4\ell})$
& $\sigma > 250$ GeV, $g_{ggX}(M_{4\ell})$ \\
\hline\hline
$\mathcal{O}_1$ & $0.005$ & $0.004$ & $0.009$ & $0.008$ \\ \hline
$\mathcal{O}_2$ & $0.099$ & $0.083$ & $0.171$ & $0.152$ \\ \hline
$\mathcal{O}_3$ & $0.206$ & $0.186$ & $0.366$ & $0.341$ \\ \hline
$\mathcal{O}_4$ & $18.2$ & $18.2$ & $4.54$ & $4.53$ \\ \hline
$\mathcal{O}_5$ & $0.023$ & $0.018$ & $0.037$ & $0.032$ \\ \hline
LO BG & $38.8$ & $13.1$ & $38.8$ & $13.1$ \\ \hline
\end{tabular}
\caption{\label{tab:xsec-table} Integrated cross sections in
  femtobarns for the $2e2\mu$
  final state without event selections for various signal processes and the LO
  irreducible background.  The signal cross sections have been
  normalized to give the SM Higgs boson on-resonance cross section.  Values
  are given both for a fixed $ggX$ coupling and assuming
  the SM evolution of this quantity with invariant mass.}
\end{table}
%%%%%%%%%%%%% End OF TABLE ################%%%%%%%%%%%%%

\subsection{Analytic Expressions for Off-Peak Cross Sections}

To gain a greater understanding of the behavior of the various cross
sections at large invariant mass, we obtain analytic expressions for
the partonic differential cross section
$\frac{d\hat{\sigma}(\hat{s})}{ d M_{Z_1} d M_{Z_2}}$.  These
expressions are valid in general, though we have suppressed the
dependence on the Higgs boson width, as our interest is in the regime where
the Higgs boson is not on-shell.
Specifically, we find that 
\begin{eqnarray}\label{sig_dM1_dM2}
\frac{d\hat{\sigma}(\hat{s})}{ d M_{Z_1} d M_{Z_2}} & = & 
g_{ggX}^2 (g_L^2 + g_R^2)^2 
\bigg( \frac{ M_{Z_1}^5 M_{Z_2}^5 \sqrt{x}}
{2^{14} 3^2 \pi^5 v^2 \hat{s}^2} \bigg)
\bigg(\frac{\hat{s}}{\hat{s}-M_X^2}\bigg)^2 
\\  \nonumber & & 
\bigg( \frac{(2 M_{Z_1} d M_{Z_1})(2  M_{Z_2} d M_{Z_2})}{
(M_{Z_1}^2 - M_Z^2)^2 + M_Z^2 \Gamma_Z^2)
(M_{Z_2}^2 - M_Z^2)^2 + M_Z^2 \Gamma_Z^2)
} \bigg) \sum_{i,j} \kappa_i \kappa_j \chi_{ij},  
\end{eqnarray}
where, using the coupling of the $Z$ to charged leptons, we have that
\begin{equation}
g_L^2 + g_R^2 = 16 \pi^2 \alpha_{EM}(\hat{s})^2\bigg(
\frac{ 2 \sin^4{\theta_W} - \sin^2{\theta_W} + 1/4}{\sin^2{\theta_W}
  \cos^2{\theta_W}} \bigg),
\end{equation}
and $x$ is defined, analogously to Refs.~\cite{Gao:2010qx, Bolognesi:2012mm}, by
\begin{equation}
x=\bigg(\frac{\hat{s}-M_{Z_1}^2 - M_{Z_2}^2}{2 M_{Z_1} M_{Z_2}}\bigg)^2-1.
\end{equation}
The expressions for the unique, non-vanishing $\chi_{ij}$ are
\begin{eqnarray}\label{os-analytic}
\chi_{11} &=& (3+x)\left(\frac{M_Z^2}{M_{Z_1} M_{Z_2}}\right)^2 , \\
\chi_{12} &=& -\frac{3}{2}\left(\frac{M_Z^2}{M_{Z_1} M_{Z_2}}\right)^2 \left(\frac{\hat s}{M_Z^2}-\frac{M_{Z_1}^2+M_{Z_2}^2}{M_Z^2}\right) , \label{c11}\\
\chi_{14} &=& (3+x) \left(\frac{M_Z^2}{M_{Z_1} M_{Z_2}}\right)^2 \left(\frac{\hat s}{M_X^2} \right), \\
\chi_{15} &=& (3+x)\left(\frac{M_Z^2}{M_{Z_1} M_{Z_2}}\right)^2 \left(\frac{M_{Z_1}^2+M_{Z_2}^2}{M_{Z}^2} \right), \\
\chi_{22} &=& 3+2 x , \\
\chi_{24} &=& -\frac{3}{2}  \left(\frac{M_Z^2}{M_{Z_1} M_{Z_2}}\right)^2 \left( \frac{\hat s}{M_X^2}\right)    \left(\frac{\hat s}{M_Z^2}-\frac{M_{Z_1}^2+M_{Z_2}^2}{M_Z^2}\right)  , \\
\chi_{25} &=& -\frac{3}{2} \left(\frac{M_Z^2}{M_{Z_1} M_{Z_2}}\right)^2 \left(\frac{M_{Z_1}^2+M_{Z_2}^2}{M_Z^2} \right)  \left(\frac{\hat s}{M_Z^2}-\frac{M_{Z_1}^2+M_{Z_2}^2}{M_Z^2}\right)   ,\\
\chi_{33} &=& 2 x, \\
\chi_{44} &=& (3+x)\left(\frac{M_Z^2}{M_{Z_1} M_{Z_2}}\right)^2 \left( \frac{\hat s}{M_X^2}\right)^2 , \\
\chi_{45} &=&  (3+x)\left(\frac{M_Z^2}{M_{Z_1} M_{Z_2}}\right)^2 \left(\frac{M_{Z_1}^2+M_{Z_2}^2}{M_Z^2} \right)\left( \frac{\hat s}{M_X^2}\right) , \\
\chi_{55} &=& (3+x)\left(\frac{M_Z^2}{M_{Z_1} M_{Z_2}}\right)^2 \left(\frac{M_{Z_1}^2+M_{Z_2}^2}{M_Z^2}\right)^2 . \label{c55}
\end{eqnarray}
We have defined these quantities such that $\chi_{ij} = \chi_{ji}$.
Note that $\chi_{i3} = 0$ for $i \ne 3$, essentially due to the parity
properties of the operators.
Eq.~(\ref{sig_dM1_dM2}) is normalized for
the $4e$ or
$4\mu$ final state (though it does not include the effects of
interference between lepton pairs; see, e.g., Ref.~\cite{Avery:2012um} 
for more discussion of this effect).  To obtain the differential cross section
for the $2e2\mu$ final state, one must multiply by two.

We now proceed to obtain expressions for the partonic cross section,
$\hat{\sigma}(\hat{s})$, by using the narrow width approximation to
integrate over $M_{Z_1}$ and $M_{Z_2}$.  The result is that 
\begin{equation}
\hat{\sigma}(\hat{s}) = 
g_{ggX}^2 
\bigg( \frac{\sqrt{1 - 4 M_Z^2 / \hat{s}} M_Z^4}
{512 \pi v^2 \hat{s}} \bigg)
\bigg(\frac{\hat{s}}{\hat{s}-M_X^2}\bigg)^2 
\sum_{i,j} \kappa_i \kappa_j \xi_{ij} (\mathrm{BR}(Z \to l^+ l^-))^2,
\end{equation}
where $\mathrm{BR}(Z \to l^+ l^-)$ gives the branching ratio for $Z$
decay to a specific lepton flavor.  As in Eq.~(\ref{os-analytic}), this expression
gives the cross section for the $4e$ or $4\mu$ final states; the value
for the $2e2\mu$ final state is greater by a factor of two.  The
$\xi_{ij}$ can be found using the expression  
\begin{equation}
\xi_{ij} = \lim_{M_{Z_{1,2}} \to~M_Z} ~\chi_{ij}.
\end{equation}
Explicitly the values of $\xi_{ij}$ are 
\begin{eqnarray}\label{os-analytic-integrated}
\xi_{11} & = & \frac{\hat{s}^2}{4 M_Z^4} - \frac{\hat s}{M_Z^2} + 3 \label{x11} \\ 
\xi_{12} & = & -\frac{3\hat{s}}{2 M_Z^2} + 3 \\
\xi_{14} & = & \bigg(\frac{\hat{s}}{M_X^2} \bigg)
\bigg( \frac{\hat{s}^2}{4 M_Z^4} - \frac{\hat s}{M_Z^2} + 3 \bigg)\\
\xi_{15} & = & 2\bigg( \frac{\hat{s}^2}{4 M_Z^4} - \frac{\hat s}{M_Z^2} + 3
\bigg) \\
\xi_{22} & = & \frac{\hat{s}^2}{2 M_Z^4} - \frac{2\hat{s}}{M_Z^2} + 3\\
\xi_{24} & = & \bigg(\frac{\hat{s}}{M_X^2} \bigg) \bigg(
-\frac{3\hat{s}}{2 M_Z^2} + 3 \bigg)\\
\xi_{25} & = & 2 \bigg( -\frac{3\hat{s}}{2 M_Z^2} + 3 \bigg) \\
\xi_{33} & = & \frac{\hat{s}^2}{2 M_Z^4} - \frac{2 \hat{s}}{M_Z^2} \\
\xi_{44} & = & \bigg( \frac{\hat{s}}{M_X^2}\bigg)^2 \bigg(
\frac{\hat{s}^2}{4 M_Z^4} - \frac{\hat s}{M_Z^2} + 3 \bigg) \\
\xi_{45} & = & 2\bigg(\frac{\hat{s}}{M_X^2} \bigg)
\bigg( \frac{\hat{s}^2}{4 M_Z^4} - \frac{\hat s}{M_Z^2} + 3 \bigg) \\
\xi_{55} & = & 4\bigg( \frac{\hat{s}^2}{4 M_Z^4} - \frac{\hat s}{M_Z^2} + 3
\bigg). \label{x55}
\end{eqnarray}
We note that many of these expressions can be obtained from the
relations
\begin{equation}
\xi_{i4} = (\hat{s}/ M_X^2) \xi_{i1}
\end{equation}
and
\begin{equation}
\xi_{i5} = 2 \xi_{i1}.
\end{equation}
As was the case for $\chi_{ij}$, $\xi_{ij} = \xi_{ji}$ and $\xi_{i3} =
0$ when $i \ne 3$.

\begin{figure}[t]
\includegraphics[width=0.47 \textwidth]{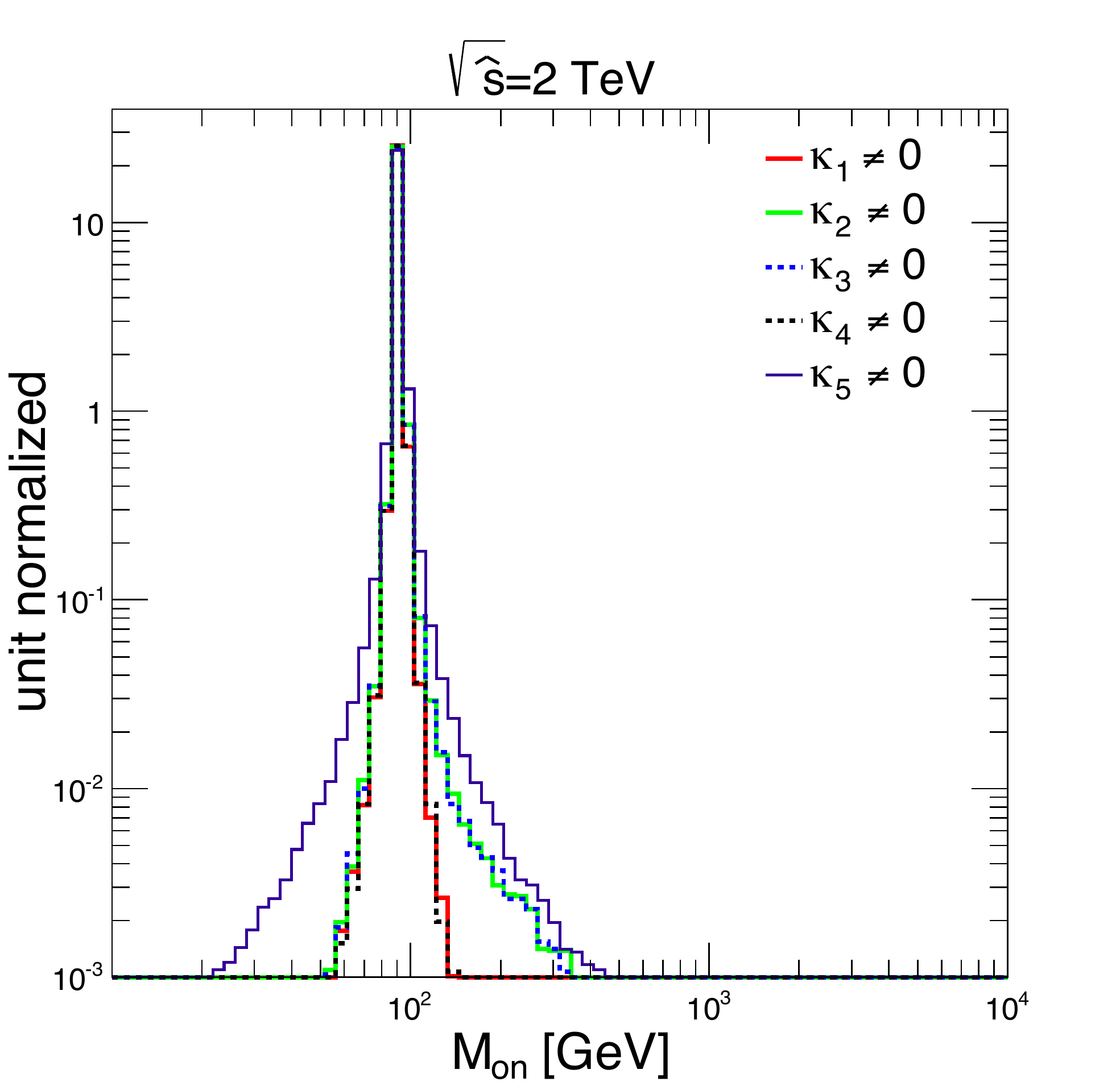} 
\includegraphics[width=0.47 \textwidth]{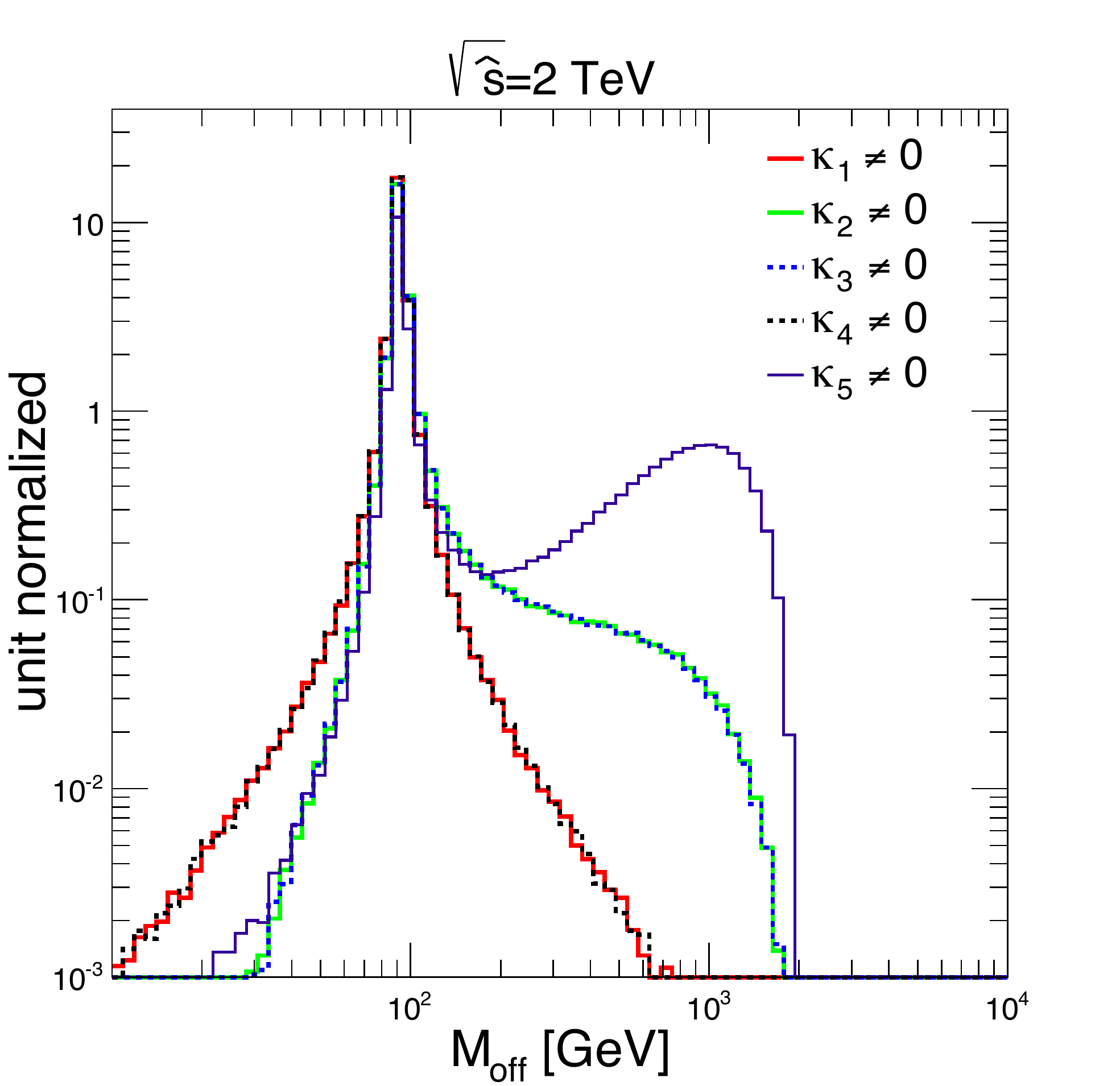} 
\caption{The distribution of $Z$ invariant mass for the $Z$ with
  invariant mass closest to $M_Z$ ($M_{\rm ON}$, left) and the $Z$
  with invariant mass furthest from $M_Z$ ($M_{\rm OFF}$, right), in
  $gg \to X \to ZZ \to 2e 2\mu$ events with $\hat{s} = 2$ TeV.  The
  curve labeled ``$\kappa_i \ne 0$'' is the distribution for which $\kappa_i$ is
  non-vanishing but $\kappa_j = 0$ for $i\ne j$; these curves have
  the same colors as the corresponding curves in Figs.~\ref{fig:2e2mu-xsec-ggX-fixed} and
  \ref{fig:2e2mu-xsec-ggX-evolve}.  We learn that a significant fraction of events from
  $\mathcal{O}_5$, and to a lesser extent $\mathcal{O}_2$, involve
  very off-shell $Z$ bosons.}\label{fig:M_ON/OFF}
\end{figure}

\begin{figure}[t]
\includegraphics[width=0.75 \textwidth]{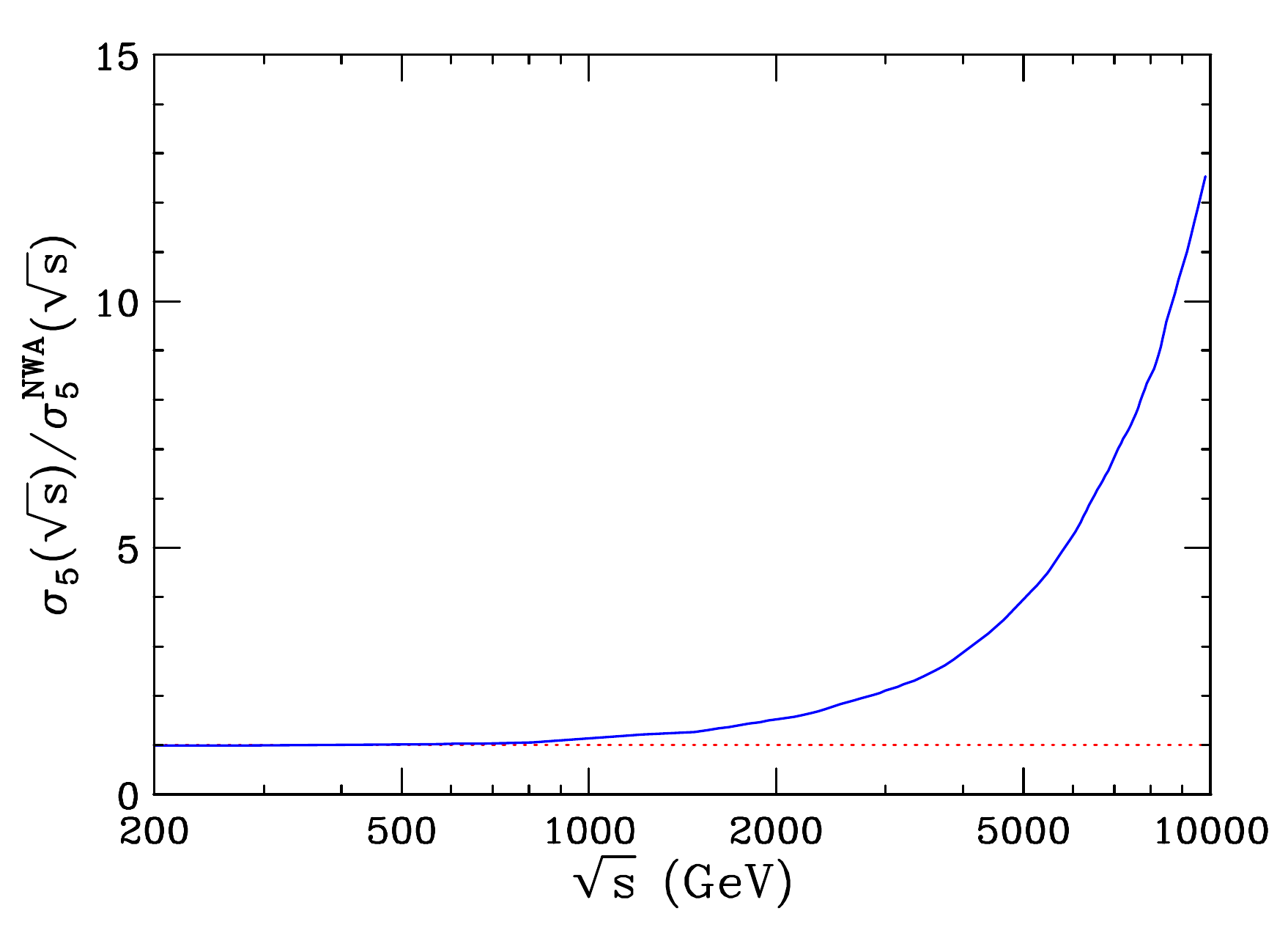} % Here is how to
                                % import EPS art
%\vspace{-36 pt}
\caption{The ratio between the actual partonic
  $gg \to ZZ^\ast \to 2e 2\mu$ cross section for pure $\mathcal{O}_5$
  couplings, and the value of this partonic cross section calculated
  in Eq.~(\ref{os-analytic-integrated}) using the narrow width
  approximation (NWA).}\label{fig:o5_ratio_plot}
\end{figure}

Some observations about the cross sections from the various operators are as follows:
\begin{itemize}
\item $\mathcal{O}_2$:  As noted above, the value of $\kappa_2$ which
  gives the SM partial width when all other couplings vanish is
  $\kappa_2 = \gamma_{22}^{-1/2}$.  (See Table~\ref{gamma table} for
  the values of the $\gamma_{ij}$.)  Thus, in the high invariant mass
  limit,
\begin{equation}
  \lim_{\sqrt{\hat{s}} \to \infty}
  \frac{\sigma_2(\sqrt{\hat{s}})}{\sigma_1(\sqrt{\hat{s}})} 
  = \frac{1}{\gamma_{22}} 
  \lim_{\sqrt{\hat{s}} \to \infty}
  \frac{\xi_{22}(\sqrt{\hat{s}})}{\xi_{11}(\sqrt{\hat{s}})}
=\frac{2}{\gamma_{22}} \approx 22.
\end{equation}
Naively, it might be surprising that $\xi_{22} / \xi_{11}$ asymptotes
to a constant value in the high $\sqrt{\hat{s}}=M_{4\ell}$ limit, as
$\mathcal{O}_2$ is built of the operators given in
(\ref{lag 2}) and (\ref{lag 3}) that are higher dimensional
than $\mathcal{O}_1$.  However, the contributions to the helicity
amplitudes from these higher dimensional operators which depend on the
highest powers of $\hat{s}$ cancel.  This cancellation is related to
the preservation of unitarity by gauge invariant operators.
\item $\mathcal{O}_3$:
Using the analogous procedure, we find that 
\begin{equation}
  \lim_{\sqrt{\hat{s}} \to \infty}
  \frac{\sigma_3(\sqrt{\hat{s}})}{\sigma_1(\sqrt{\hat{s}})} 
=\frac{2}{\gamma_{33}} \approx 53.
\end{equation}
Again, the fact that the highest power of $\hat{s}$ in $\xi_{33}$ is
two is related to the gauge invariance of the $\mathcal{O}_3$ operator.
\item $\mathcal{O}_4$:
Here there is a dramatic enhancement of the cross section at high
energies as 
\begin{equation}
  \lim_{\sqrt{\hat{s}} \to \infty}
  \frac{\sigma_4(\sqrt{\hat{s}})}{\sigma_1(\sqrt{\hat{s}})} 
  = \frac{\hat{s}^2}{M_X^4}.
\end{equation}
The tendency for amplitudes associated with this operator to grow with
energy leads to issues with unitarity, as we will discuss in more
detail in Subsection~\ref{The U}.
\item $\mathcal{O}_5$:
If we use the expressions for $\xi_{11}$ and $\xi_{55}$ in
Eq.~(\ref{x11}) and Eq.~(\ref{x55}), then we would obtain
\begin{equation}
  \lim_{\sqrt{\hat{s}} \to \infty}
  \frac{\sigma_5(\sqrt{\hat{s}})}{\sigma_1(\sqrt{\hat{s}})} 
  = \frac{4}{\gamma_{55}} \approx 4.
\end{equation}
However, following Eq.~(\ref{c11}) and Eq.~(\ref{c55}),
we note that
\begin{equation}
\frac{\chi_{55}}{\chi_{11}} = \bigg(\frac{M_{Z_1}^2 + M_{Z_2}^2}{M_Z^2}\bigg)^2.
\end{equation}
The extra powers of $M_{Z_1}$ and $M_{Z_2}$ in $\chi_{55}$ mean that
the narrow width approximation (NWA), which was used in obtaining the
$\xi_{ij}$ from the $\chi_{ij}$, breaks down, leading to an
enhancement of the cross section at high invariant mass from events
with very off-shell $Z$ bosons.  
The prevalence of events with very off-shell (high invariant mass) $Z$
bosons can be seen in Fig.~\ref{fig:M_ON/OFF};
the enhancement of the partonic cross
section as a function of $\hat{s}$ is shown in
Fig.~\ref{fig:o5_ratio_plot}.
The enhancement in cross section versus the NWA expectation for 
$\mathcal{O}_5$ might seem to
promise an increase in sensitivity.
However, the interference
between signal events with large, off-shell $M_{Z_1}$ and $M_{Z_2}$
and the continuum $gg \to ZZ$ will be quite small, and the total cross
section for such events is small for LHC purposes. 
Perhaps the situation will be somewhat more optimistic at a $100$ TeV collider.
\end{itemize}
%%%%%%%%%%%%% Beginning OF FIGURE ################%%%%%%%%%%%%%
%%%%%%%%%%%%% End OF FIGURE %%%%%%%%%%%%%%%%%%%%%%%%%%%%%%%%%
%%%%%%%%%%%%% Beginning OF FIGURE ################%%%%%%%%%%%%%
%%%%%%%%%%%%% End OF FIGURE %%%%%%%%%%%%%%%%%%%%%%%%%%%%%%%%%

\subsection{Unitarity Bounds on $\mathcal{O}_4$}\label{The U}

\begin{figure}[th]
\includegraphics[width=0.75 \textwidth]{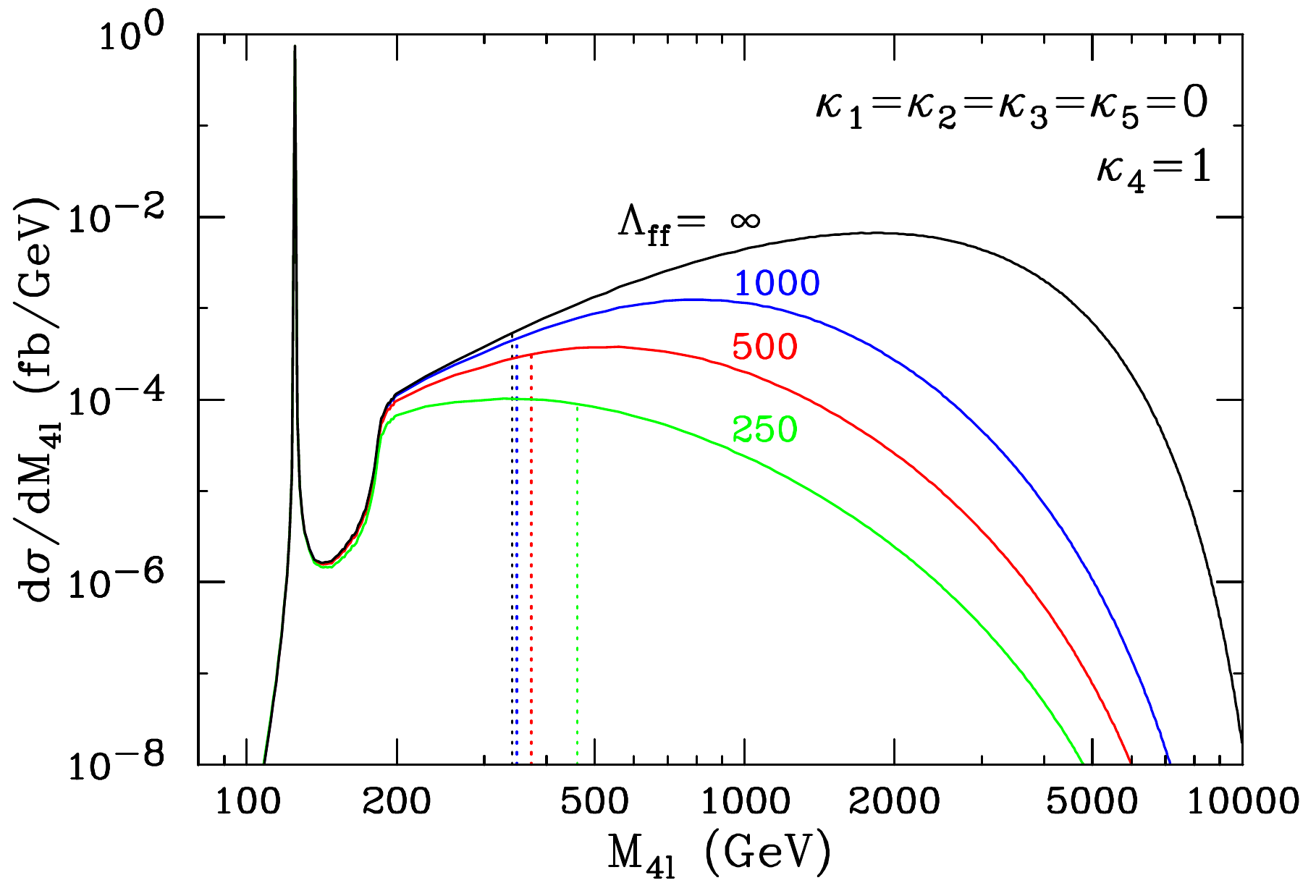} % Here is how to
                                % import EPS art
%\vspace{-36 pt}
\caption{The differential cross-section $d\sigma/dM_{4\ell}$ in fb/GeV
for pure $\mathcal{O}_4$ operator with $\kappa_4=1$, for several choices of the 
form-factor scale $\Lambda_{ff}$: $\infty$ (black line),
$1$ TeV (blue line), $500$ GeV (red line), and $250$ GeV (green line).
The vertical dotted lines denote the scale $\Lambda$ at which
unitarity violation occurs in each case. The cross section considered
(and listed in Table~\ref{tab:lambda}) in each form factor scenario is
that found by integrating under the relevant cross section curve up to
the relevant dotted line. 
\label{fig:ff}
}
\end{figure}

A striking feature in Figs.~\ref{fig:2e2mu-xsec-ggX-fixed} and
~\ref{fig:2e2mu-xsec-ggX-evolve}, as well as in the integrated cross
sections in Table~\ref{tab:xsec-table}, is the rapidly growing cross
section from the $\mathcal{O}_4$ operator.  
However, the growth in the strength of this operator with invariant
mass will lead to amplitudes which violate partial wave unitarity at
some mass scale $\Lambda$~\cite{Lee:1977eg, Dobado:1999xb, 
Dahiya:2013uba,Delgado:2013hxa}.  
Three approaches to this issue are, in increasing order of
conservatism,
\begin{enumerate}
\item Ignore unitarity and set limits using the entire
  predicted off-shell cross section.
\item Use form factors, e.g., as in Ref.~\cite{Chatrchyan:2012sga,
    PatriciaRebelloTelesfortheCMS:2013sra}, 
  that prevent the amplitude from violating unitarity or 
  at least increase the mass scale at which unitarity is violated.
\item Consider only cross sections for invariant masses
  less than $\Lambda$. 
\end{enumerate}
We demonstrate how one obtains the predicted off-shell cross section
in each of these approaches in Fig.~\ref{fig:ff}.  We note that options 2
and 3 both require a study of the unitarity bounds on $\kappa_4$.

%%%%%%%%%%%%% Beginning OF FIGURE ################%%%%%%%%%%%%%
%%%%%%%%%%%%% End OF FIGURE %%%%%%%%%%%%%%%%%%%%%%%%%%%%%%%%%

Unitarity violating behavior can be probed in a variety of
channels~\cite{Lee:1977eg, Gunion:1989we, Dobado:1999xb,
  Dahiya:2013uba, Delgado:2013hxa}.
As we have specified only an effective theory of
the $XZZ$ coupling, we will look only at $Z_L Z_L \to Z_L Z_L$
scattering, as our study of this process requires no assumptions
beyond the Lagrangian presented above in Eq.~(\ref{Lag}).  
However,
in the well-motivated limit where $SU(2)$ symmetry is
spontaneously broken, we would expect the $XWW$ couplings to be related,
allowing the study of additional channels.  

Longitudinal $ZZ$ scattering involves three diagrams --- $s$, $t$, and
$u$-channel scalar exchange.  In the limit where $M_Z$ (and, of course,
$\Gamma_H$) can be neglected, the contribution to the $J=0$ partial
wave when $\kappa_4$ is non-zero, $\kappa_2 = \kappa_3 = \kappa_5 =
0$, and $\kappa_1 = 1 - \kappa_4$ (to ensure that one obtains the SM
value of the partial width), is
\begin{eqnarray}\label{eq:a0}
a_0(s) = \bigg( \frac{M_X^2}{32 \pi v^2} \bigg) \bigg[
\frac{(s/M_X^2)^2}{6} \bigg( (10 - 3 s/M_X^2) \kappa_4^2 - 20 \kappa_4
\bigg) - & \\
\bigg(3 + \frac{M_X^2}{s - M_X^2} - \frac{2 M_X^2}{s} \log{(1
+ \frac{s}{M_X^2})}\bigg)\bigg],
\end{eqnarray}
where we have included the factor of $1/2$ from the normalization of
$ZZ$ in our expression for $a_0$~\cite{Lee:1977eg, Gunion:1989we}.  

Even for relatively moderate values of $\sqrt{s}$, $a_0(s)$ is
dominated by the $\kappa_4^2$ and $\kappa_4$ terms.  Clearly at very
high values of s, we have
\begin{eqnarray}\label{eq: a0 high}
a_0(s) \sim \frac{- s^3 \kappa_4^2}{64\pi M_X^4 v^2}.
\end{eqnarray}
Thus, the $s$-dependence of this quantity is three powers greater than
for its SM analogue, which asymptotes to a constant value.  
Two of these three additional powers are due to the
$s$ dependence of the $\mathcal{O}_4$ vertex,
while the third is due to a
failure of the unitarity-preserving cancellation between amplitudes
that cause the SM amplitude to approach a constant at high energies.

Using expression~(\ref{eq: a0 high}), an approximate unitarity
bound found by setting $|\mathrm{Re}~a_0(\Lambda^2)| = 1/2$
(cf. Ref.~\cite{Gunion:1989we}) is
\begin{eqnarray}
\Lambda = (32 \pi M_X^4 v^2)^{1/6} |\kappa_4|^{-1/3}.
\end{eqnarray}
As $(32 \pi M_X^4 v^2)^{1/6} \approx 340~$GeV, it is clear that we
cannot neglect the other terms in $a_0(s)$.  

\begin{figure}[t]
\includegraphics[width=0.47 \textwidth]{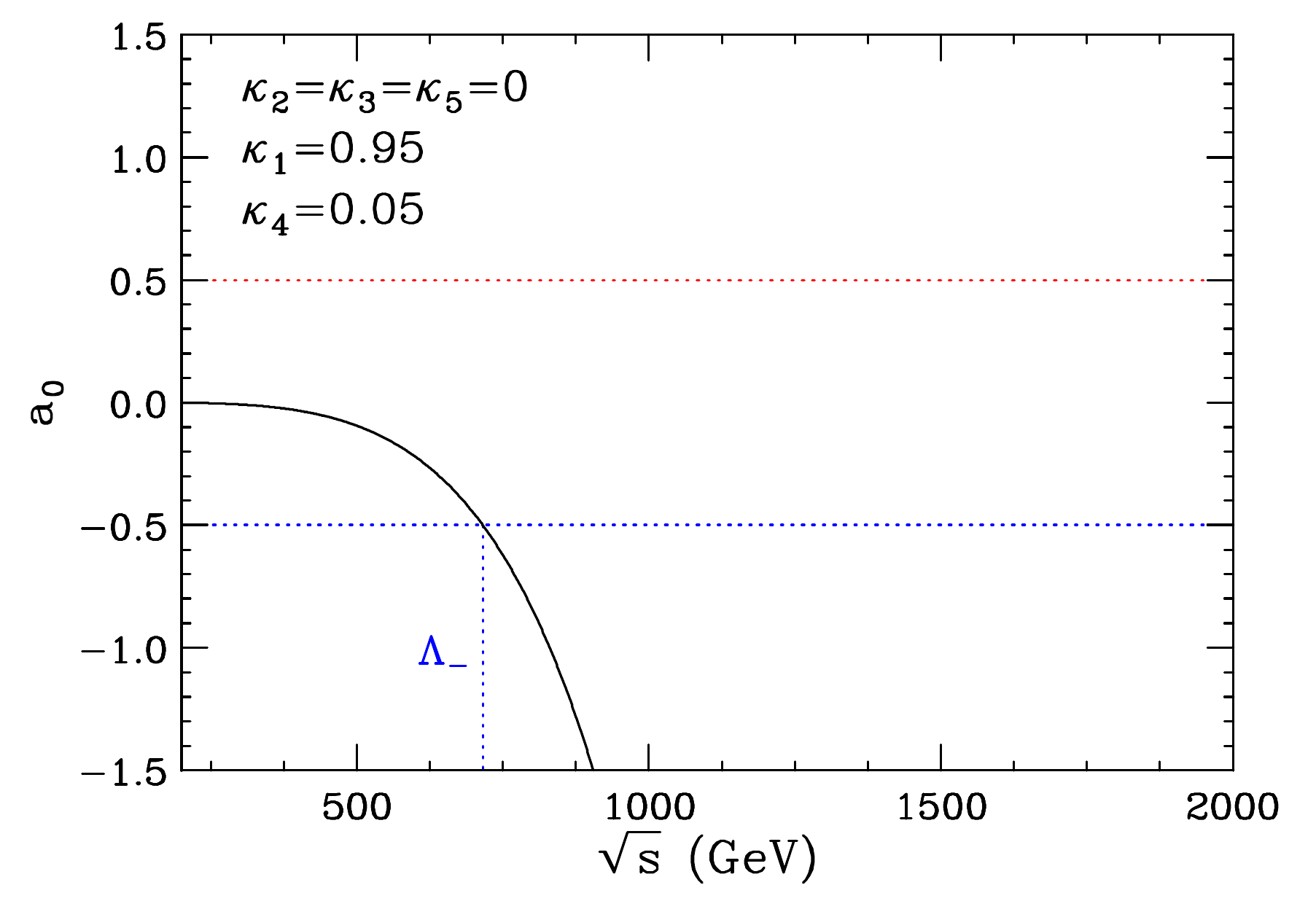} 
\includegraphics[width=0.47 \textwidth]{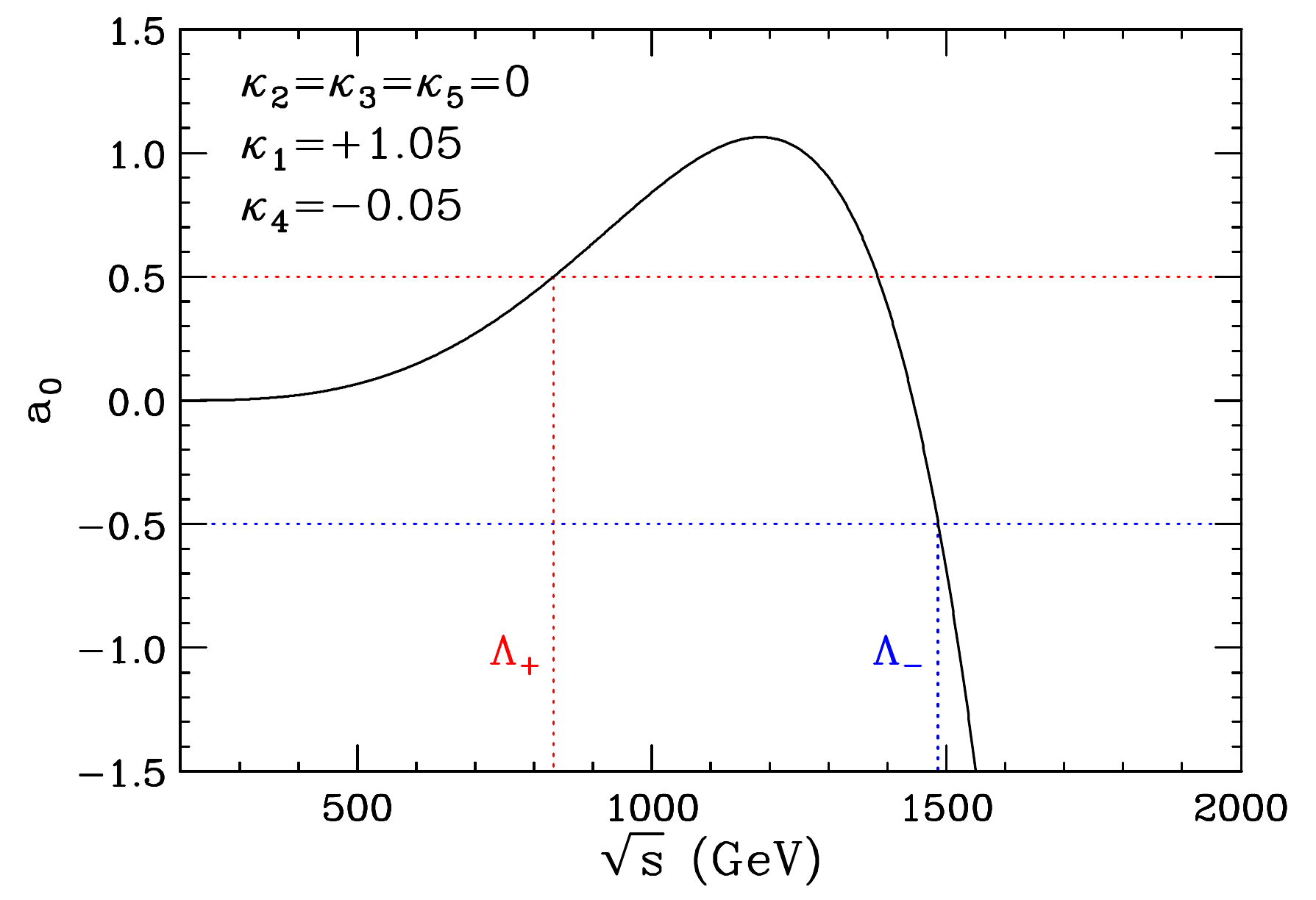} \\
\includegraphics[width=0.47 \textwidth]{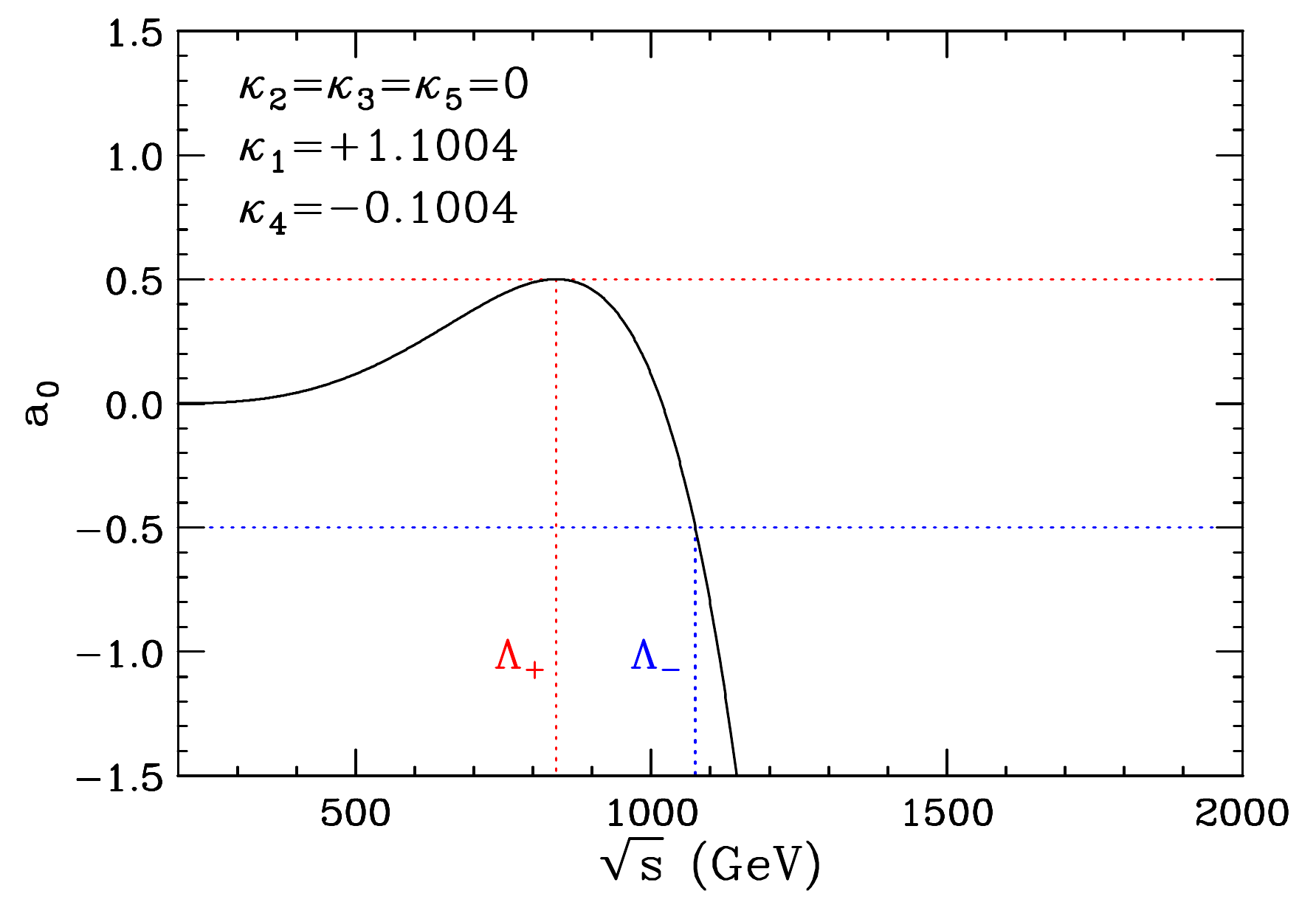} 
\includegraphics[width=0.47 \textwidth]{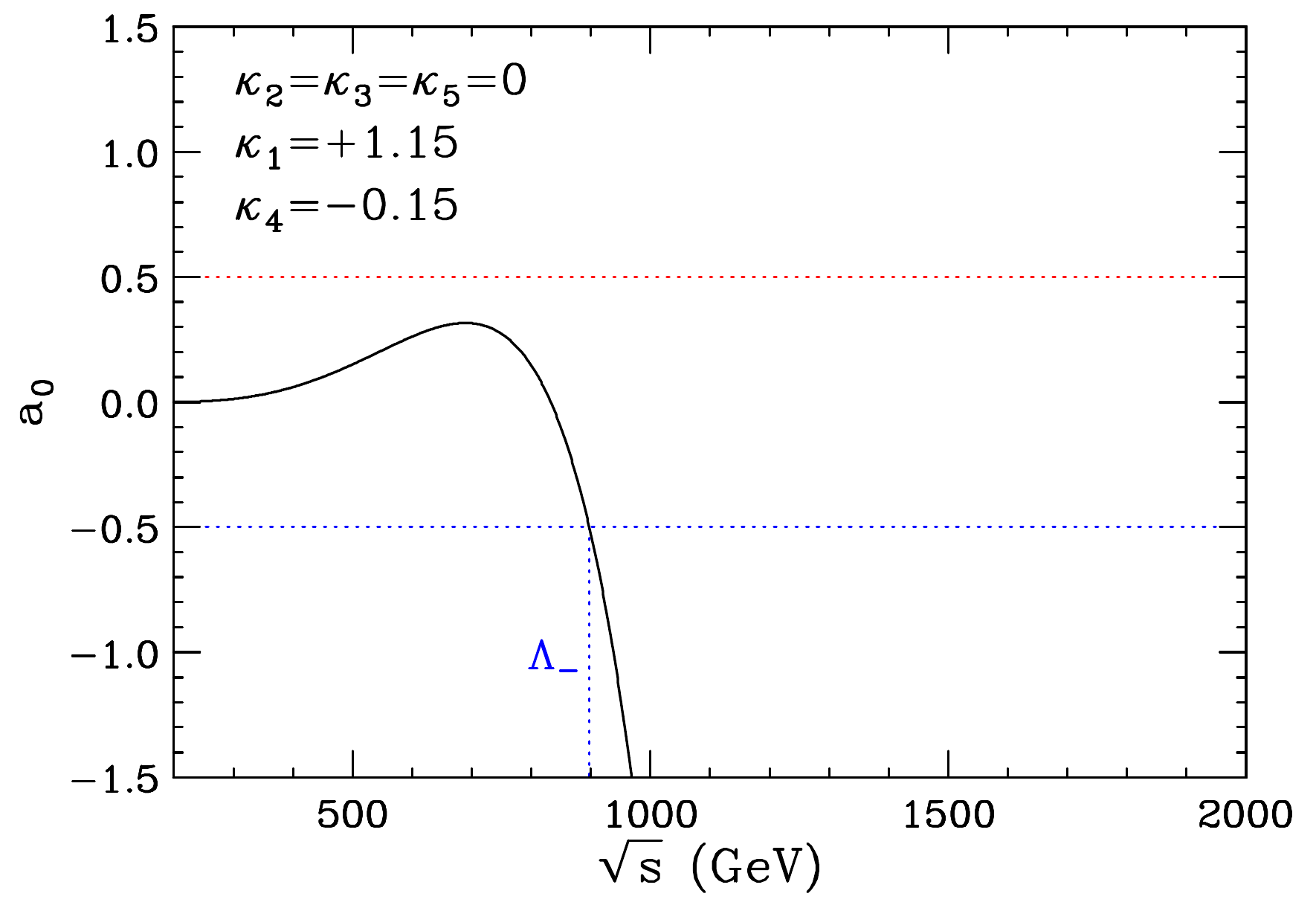} 
%\vspace{-36 pt}
\caption{\label{fig:a0}
The contribution to the $J=0$ partial wave for the $Z_L Z_L \to Z_L
Z_L$ scattering amplitude, $a_0(s)$, shown as a function of $\sqrt{s}$
for several different values of $\kappa_4$ to illustrate the
qualitative differences in the behavior of this function for different
values of $\kappa_4$.  In all cases $\kappa_1 =
1 - \kappa_4$ to ensure that the point considered gives the SM partial
width.}
\end{figure}

Considering now the entire
term proportional to $\kappa_4^2$ in Eq.~(\ref{eq:a0}), we note that
for either sign of $\kappa_4$, this term is positive for $s <
\sqrt{10/3} M_H \approx 230$ GeV and negative thereafter.  The term
linear in $\kappa_4$ gives a negative contribution when $\kappa_4$ is
positive and a positive contribution when $\kappa_4$ is negative.
Thus, if $\kappa_4 > 0$ (and if the unitarity bound $\Lambda$ is greater
than $\approx 230$ GeV) then the unitarity bound will occur when
$a_0(s)$ becomes sufficiently \emph{negative}, i.e. when
$a_0(\Lambda^2) = -1/2$.
However, if $\kappa_4 < 0$, then $a_0(s)$ will be positive up to some
scale $> 230$ GeV, and possibly much greater.  
At sufficiently high
values of $s$, the curve must turn negative and approach
expression~(\ref{eq: a0 high}) at high energies.  So the minimal (and
hence the physically interesting) scale at which partial wave
unitarity is violated may occur for either positive or negative values
of $a_0(s)$.  
Defining $\Lambda_\pm$ to be the lowest value of
$\sqrt{s}$ for which $a_0(s) = \pm 1/2$, we demonstrate the behavior
of $a_0(s)$ for various choices of $\kappa_4$ in Fig.~\ref{fig:a0}.
We find numerically that when $\Lambda_+$ exists, it is often
approximately equal to $\Lambda_{\rm{linear}}$, the value of the
unitarity bound if $a_0(s)$ contained only the term linear in
$\kappa_4$, while $\Lambda_-$ is generally closer to
$\Lambda_{\rm{quad}}$, the value of the unitarity bound if $a_0(s)$
contained only the term quadratic in $\kappa_4$.

%%%%%%%%%%%%% Beginning OF FIGURE ################%%%%%%%%%%%%%
%%%%%%%%%%%%% End OF FIGURE %%%%%%%%%%%%%%%%%%%%%%%%%%%%%%%%%

\begin{figure}[th]
\includegraphics[width=0.75 \textwidth]{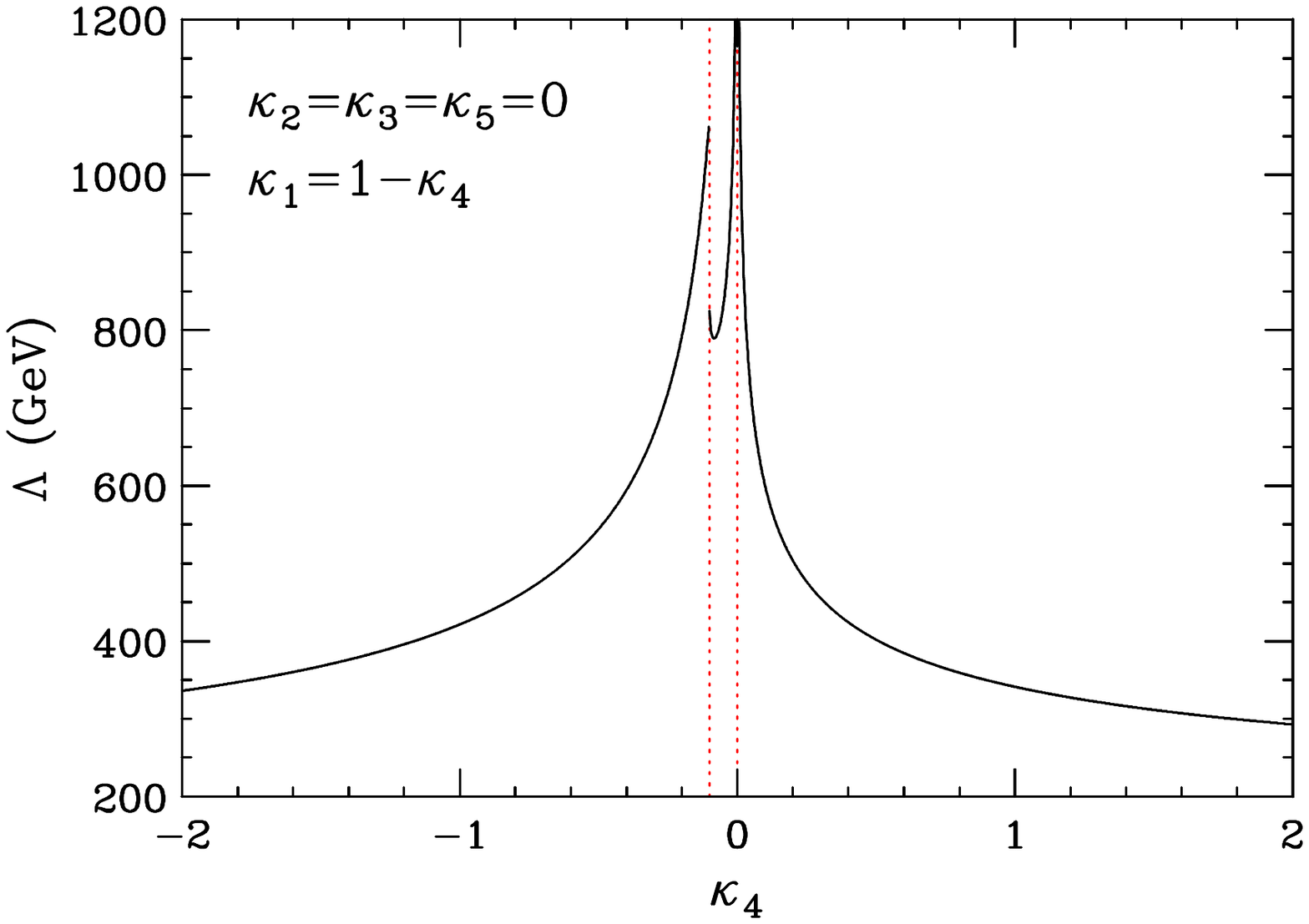} % Here is how to
                                % import EPS art
%\vspace{-24 pt}
\caption{
The scale, $\Lambda$, at which partial wave unitarity is violated $\kappa_4$,
for an admixture of $\kappa_1$ and $\kappa_4$ couplings.  As above,
$\kappa_1 = 1 - \kappa_4$ so that the one obtains the SM value for the
$X \to 4\ell$ partial width.
\label{fig:unitarity}
}
\end{figure}

Based on this understanding of the behavior of $a_0(s)$ for various values
of $\kappa_4$, a unitarity bound as a function of
$\kappa_4$ can be determined, which is shown in Fig.~\ref{fig:unitarity}. 
We note that the transition from the region where the unitarity bound
is $\approx \Lambda_{\rm{quad}}$ to the region where the unitarity
bound is $\approx \Lambda_{\rm{linear}}$ at $\kappa_4 =
\kappa_{4,~\rm{special}}$ provides a ``first order transition''.  This
is because for values of $\kappa_4$ slightly greater than the values
at this point, $|a_0(s)| = 1/2$ for both positive and negative values
of $a_0(s)$, while for values slightly less than the values at this
point, $|a_0(s)| = 1/2$ only occurs when $a_0(s) = -1/2$. At this
point, $\kappa_{4,~\rm{special}} \approx -0.1004$, the maximum of $a_0(s)$ is equal to
$1/2$, as may be seen in the bottom left plot in Fig.~\ref{fig:a0}.

%%%%%%%%%%%%% Beginning OF FIGURE ################%%%%%%%%%%%%%
%%%%%%%%%%%%% End OF FIGURE %%%%%%%%%%%%%%%%%%%%%%%%%%%%%%%%%

In a conservative approach to taming the high-energy behavior
of $\mathcal{O}_4$,
we only consider events with $M_{4\ell} <
\Lambda(\kappa_4)$ when excluding a particular value of $\kappa_4$.
A less conservative approach is to consider a
``form factor'' scenario in $\kappa_4$ depends on
$s$ as follows:
\begin{equation}\label{form_factor}
\kappa_4 \to  \frac{1 + M_X^2/\Lambda_{ff}^2}
{1 + s/\Lambda_{ff}^2} \times \kappa_4.
\end{equation}
The expression in the numerator is only a normalization used 
to ensure
that $\kappa_4(M_X)$ is unchanged by the transformation.
As noted above, the high energy behavior of $a_0(s)$, in the absence
of form factors, goes as the third power of $s$.  Hence the
transformation in Eq.~(\ref{form_factor}) does not fully unitarize $Z_L
Z_L$ scattering.  Therefore, in employing this procedure, we consider
only the cross section integrated up to the unitarity bound found when
the coupling is modified as in Eq.~(\ref{form_factor}).  We also modify
the $gg \to X \to ZZ^\ast \to 4\ell$ cross section, accordingly, as
can be seen in Fig.~\ref{fig:ff}.

Cross sections obtained from this procedure are shown in
Table~\ref{tab:lambda} for several choices of the form factor scale
$\Lambda_{ff}$.  We have also included the cross section found from 
the unitarity bounds in the case where we do not modify $\kappa_4$;
this corresponds to the $\Lambda_{ff} \to \infty$ limit.
We note that these cross sections are quite modest, especially
compared with the value for $\sigma_4$ shown in
Table~\ref{tab:xsec-table} above ($18.2$/$4.54$ fb in the fixed/evolving
$ggX$ coupling scenario).  

However, the off-shell cross section for a pure $\mathcal{O}_4$
coupling is still significantly larger than the SM off-shell cross
section ($5$/$9$ ab in the fixed/evolving $ggX$ coupling scenario, 
as given in Table~\ref{tab:xsec-table}).
 As it was suggested in Ref.~\cite{Campbell:2013una} that
the LHC may be sensitive ultimately to an off-shell cross section $5$ to
$10$ times greater than the SM value, 
there is reason to hope
that one can discriminate between $\mathcal{O}_1$ and $\mathcal{O}_4$,
even when taking unitarity into account in a conservative manner.

%%%%%%%%%%%%% Beginning OF TABLE ################%%%%%%%%%%%%%
\begin{table}
\begin{tabular}{| c || c || c | c |}
\hline
$\Lambda_{ff}$ &  $\Lambda$   &   $\sigma > M_X$, all $M_{4\ell}$    &   $\sigma > M_X$, for $M_{4\ell}\le\Lambda$    \\
 (GeV)               &   (GeV)           &        (fb)                                          &     (fb)  \\
\hline
$\infty$              &  341.3   & 18.205 (4.544)	  &   0.044 (0.065)   \\
1000   	          &  349.2   &  1.526 (1.435)	  &   0.043 (0.065)   \\
500    	 		     &  373.0   &  0.333 (0.472)	  &   0.038 (0.065)   \\
250      		     & 461.8   &  	0.064 (0.107)      &  0.026 (0.053)   \\	
\hline
\end{tabular}
\caption{\label{tab:lambda} Integrated cross sections in femtobarns (at leading order) for the $2e2\mu$
  final state without event selections and for the case of a pure $\mathcal{O}_4$ operator, 
  with different values of the form factor scale $\Lambda_{ff}$.  
  The signal cross sections have been normalized to give the SM Higgs boson on-resonance cross section.  The first values
  are obtained with a fixed $ggX$ coupling, while the values in parentheses assume
  the SM evolution of this quantity with invariant mass.
  The second column shows the scale $\Lambda$ of unitarity violation.
  The results in the third (fourth) column are obtained after integrating over the whole allowed range for $M_{4\ell}$ 
  (only up to $M_{4\ell}\le \Lambda$).}
\end{table}
%%%%%%%%%%%%% End OF TABLE ################%%%%%%%%%%%%%

%%%%%%%%%%%%% Beginning OF FIGURE ################%%%%%%%%%%%%%
\begin{figure}[th]
\includegraphics[width=0.75 \textwidth]{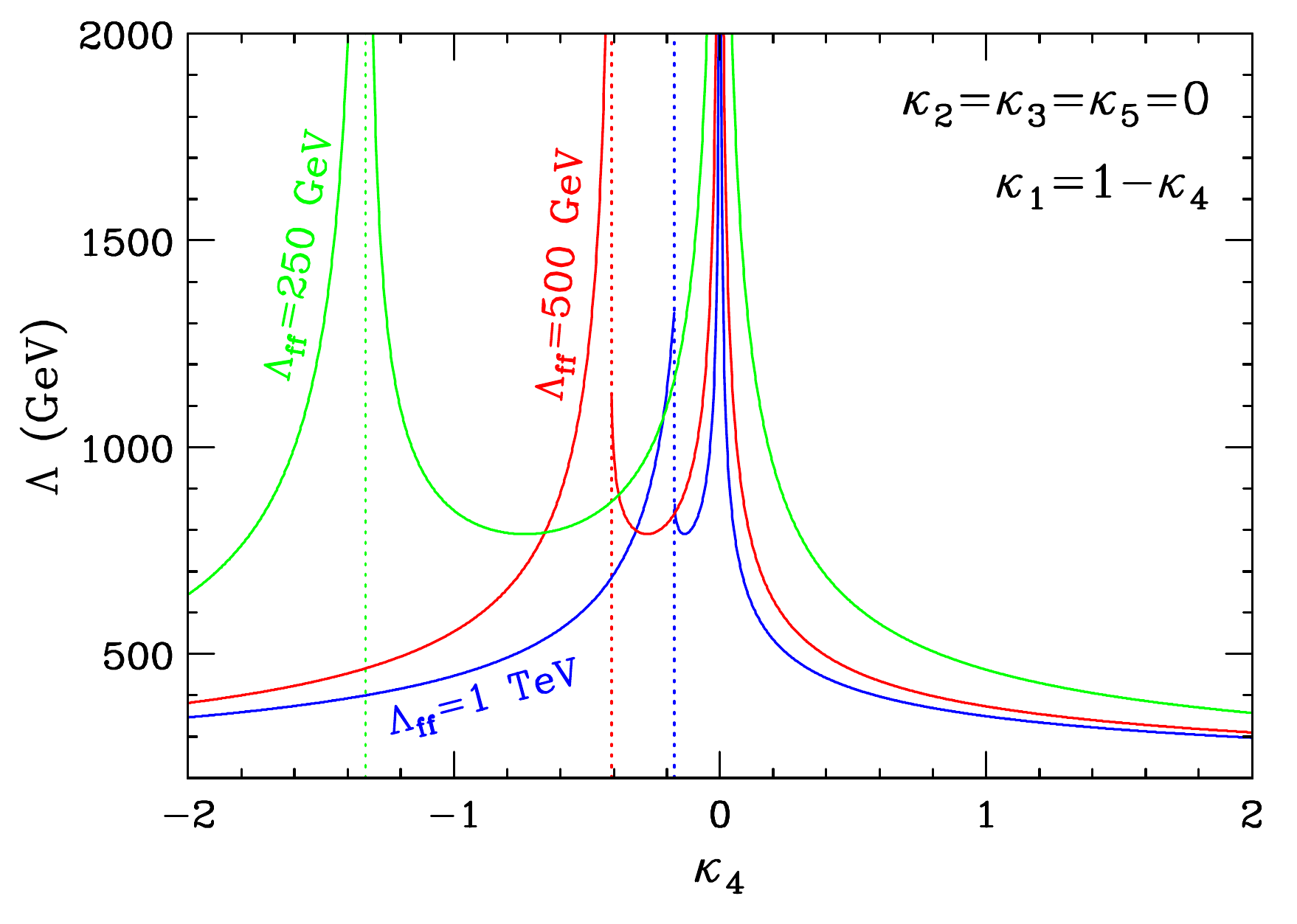} % Here is how to
                                % import EPS art
%\vspace{-24 pt}
\caption{The same as Fig.~\ref{fig:unitarity}, but in the presence of a form factor with
$\Lambda_{ff}=1$ TeV (blue),
$\Lambda_{ff}=500$ GeV (red), and
$\Lambda_{ff}=250$ GeV (green).
\label{fig:unitarity2}
}
\end{figure}
%%%%%%%%%%%%% End OF FIGURE %%%%%%%%%%%%%%%%%%%%%%%%%%%%%%%%%

\section{On-Shell Phenomenology of $XZZ$ Operators}\label{on-shell}

Now that we have shown the importance of the off-shell ($M_{4\ell} \gg
M_X$) four-lepton cross section for probing $XZZ$ couplings, we
proceed to make a few remarks about the relevant on-shell
phenomenology, focusing, in particular, on probing $\mathcal{O}_5$
couplings.  We note that interference with continuum $gg \to ZZ$ is
less important here than in the off-shell case considered
above due to one of the $Z$ bosons necessarily being off-shell.

\subsection{Distinguishing $\mathcal{O}_5$ On-Peak}

\begin{figure}[th]
\includegraphics[width=0.75 \textwidth]{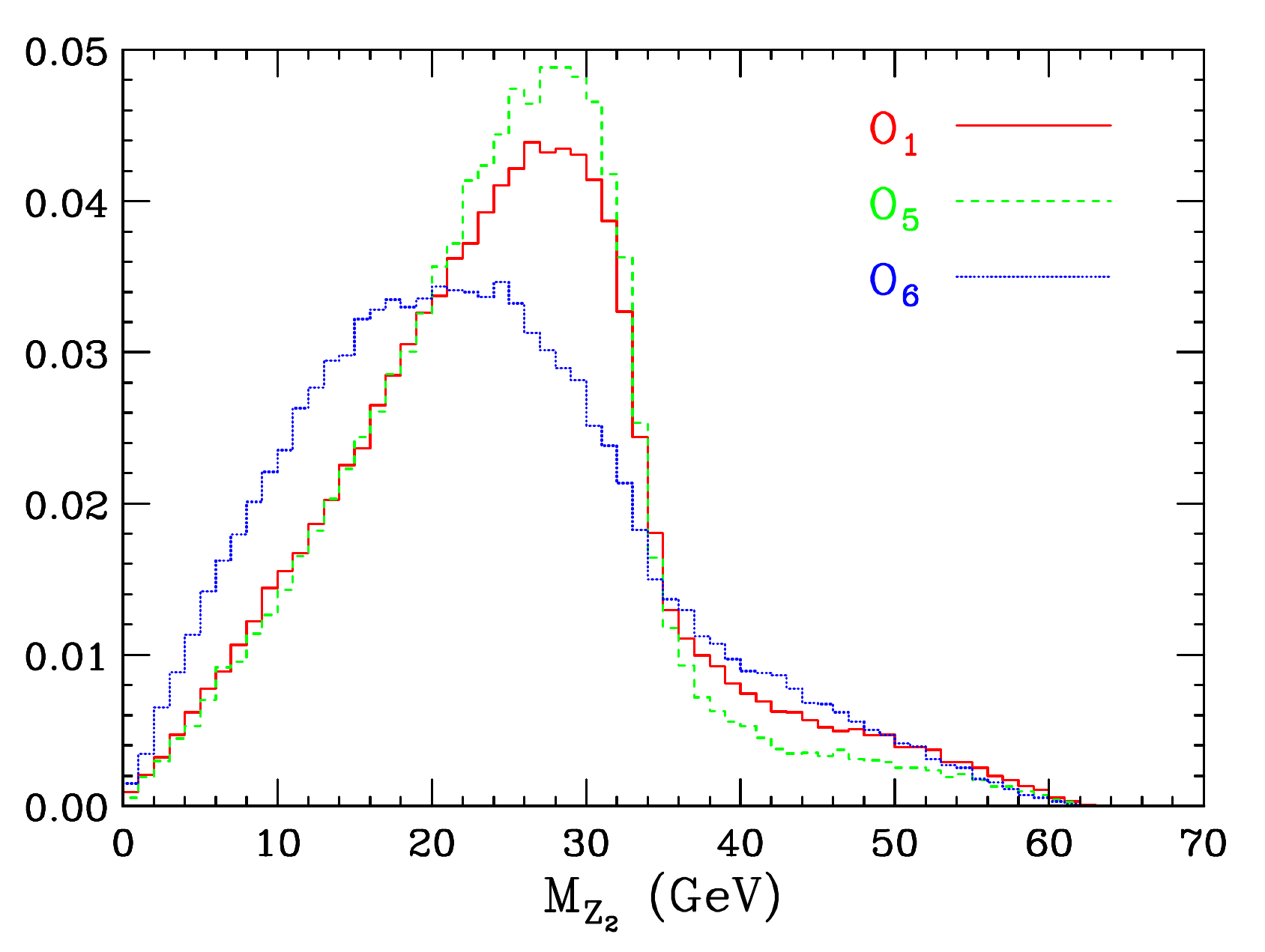} % Here is how to
                                % import EPS art
%\vspace{-36 pt}
\label{fig:k5_m2}
\caption{
This figure shows the invariant mass distribution for the
reconstructed $Z$ with lower invariant mass ($M_{Z_2}$) for pure
$\mathcal{O}_1$ (tree-level SM) couplings (red solid line), 
pure $\mathcal{O}_5$ couplings  (green dashed line), 
and pure $\mathcal{O}_6$ couplings (blue dotted line). 
}
\end{figure}

Let $\mathcal{A}_{1(5)}$ refer to the amplitude for a particular
kinematic configuration due to $\mathcal{O}_{1(5)}$.
Then
\begin{equation}
\mathcal{A}_5 = \frac{M_{Z_1}^2 + M_{Z_2}^2}{M_Z^2} \mathcal{A}_1,
\end{equation}
as alluded to above.
Thus, when $M_{4\ell} \approx M_X$, the dependence of the amplitude
on $M_{Z_1}^2 + M_{Z_2}^2$ will affect the angular and invariant distributions
of the four leptons, in particular the $M_{Z_2}$ distribution.
To demonstrate this, we compare the $M_{Z_2}$ distribution due to pure
$\mathcal{O}_1$, $\mathcal{O}_5$, and $\mathcal{O}_6$ couplings in
Fig.~\ref{fig:k5_m2}.
%%%%%%%%%%%%% Beginning OF FIGURE ################%%%%%%%%%%%%%
%%%%%%%%%%%%% End OF FIGURE %%%%%%%%%%%%%%%%%%%%%%%%%%%%%%%%%
We note that while there is a discernible difference between the SM
$\mathcal{O}_1$ distribution and the $\mathcal{O}_5$ distribution,
this difference is relatively subtle, which suggests that it may be
somewhat challenging to discover or constrain $\kappa_5$ couplings at
the LHC.  We give some idea of the extent to which this is true in the
next subsection.

\subsection{Quantifying Sensitivity to Non-SM Couplings}
The optimal discrimination between two hypotheses is that obtained
using the likelihood as the test statistic\cite{Neyman:Pearson}.  With this in
mind, we can quantify the maximum possible sensitivity for the
exclusion of non-SM Higgs boson couplings to $Z$ bosons by determining the average
value of the log likelihood ratio using SM events, namely 
\begin{equation}\label{delta log L}
\langle\Delta \log{\mathcal{L}}\rangle_{SM} =   
\left\langle \log{\bigg[ \bigg( \frac{\sigma_1}{\sigma_{\{\kappa_i\}}} \bigg)} \bigg(
  \frac{d \sigma_{\{\kappa_i\}}}{d\mathbf{x}} \bigg/
\frac{d \sigma_1}{d\mathbf{x}} \bigg)\bigg] \right\rangle_{SM}.
\end{equation}
We determine this quantity for the pure operator couplings
$\mathcal{O}_2$, $\mathcal{O}_3$, $\mathcal{O}_5$, and
$\mathcal{O}_6$; the results are shown in Table~\ref{tab:sensitivity}.   
We use only  events with $M_{4\ell} = M_X = 125$ GeV (the off-shell
cross section is of course small for the SM in any case), hence
$\langle\Delta \log{\mathcal{L}}\rangle_{SM}$ is $0$ for
$\mathcal{O}_4$, as this operator is identical to the SM
$\mathcal{O}_1$ operator when $M_{4\ell} = M_X$.

In the limit of large statistics, twice the log likelihood ratio is equivalent to
the difference in the $\chi^2$ value of the two hypotheses fit to data.
Thus, e.g. $3^2 / (2\langle\Delta \log{\mathcal{L}}\rangle_{SM})$ gives
an approximation, valid in the limit of sufficient events, for the
expected number of events required to obtain a $3\sigma$ limit on the
given pure couplings, assuming the tree level SM is the true theory.
This number will undershoot the true value, as we are taking into
account neither the irreducible SM background nor detector effects.
However, the result is reasonable.  The CMS analysis was able to rule out
a pure $\mathcal{O}_3$ coupling at slightly greater than $3\sigma$ 
and $\mathcal{O}_2$ at slightly less than $2\sigma$ with
 $\sim 20$ signal events~\cite{Chatrchyan:2013mxa}.
This suggests that the 
number of events needed to obtain a given sensitivity in
Table~\ref{tab:sensitivity} are smaller than the number of events
actually needed in an experiment by factors of $2 - 5$.  (ATLAS
reported a slightly less than $3\sigma$ exclusion of $\mathcal{O}_3$
in Ref.~\cite{1523699}; they did not report a limit on
$\mathcal{O}_2$.)

We note that the $\mathcal{O}_5$ operator, as expected, is harder to
distinguish from the SM than the $\mathcal{O}_2$ or $\mathcal{O}_3$
operators.  This justifies the postponement of the measurement of this
operator, as was suggested in Ref.~\cite{Gainer:2013rxa}.  
Assuming
the scaling between the theoretical optimum value in
Table~\ref{tab:sensitivity} and the actual number of events needed by
an experiment for a given sensitivity holds, then $100 - 200$
fb$^{-1}$ of 13 TeV running at LHC should conclusively rule out a pure
$\mathcal{O}_5$ coupling, if the Higgs boson is truly SM-like.

%%%%%%%%%%%%% Beginning OF TABLE ################%%%%%%%%%%%%%
\begin{table}
\begin{tabular}{| l | c | c | c | c | c | c |}
\hline
~ & $\mathcal{O}_1$ & $\mathcal{O}_2$ & $\mathcal{O}_3$ &
$\mathcal{O}_4$ & $\mathcal{O}_5$ & $\mathcal{O}_6$ \\ \hline
$2 \langle \Delta \log{\mathcal{L}}\rangle_{SM}$ & $0$ & ~$-0.747$~ & ~$-1.017$~ & $0$ &
  ~$-0.178$~ & ~$-0.503$~ \\ \hline
Events for $3\sigma$ Limit & ~\line(1,0){32}~ & $12.0$ & $8.85$ &
~\line(1,0){32}~ & $50.5$ & $17.9$ \\ \hline \hline
\end{tabular}
\caption{
This table gives twice the difference between the average log likelihood
obtained assuming pure couplings and 
the average log likelihood obtained assuming SM ($\mathcal{O}_1$) as
evaluated for events generated under the SM ($\mathcal{O}_1$) hypothesis.
This value is then 
used to provide an optimistic estimate of the number of
events required for a $3\sigma$ exclusion of the specified coupling.
\label{tab:sensitivity}
 }
\end{table}
%%%%%%%%%%%%% End OF TABLE ################%%%%%%%%%%%%%

\section{Conclusions}\label{Conclusions}

In this paper we have extended the framework for $XZZ$ coupling
measurements in the four-lepton final state presented in
Ref.~\cite{Gainer:2013rxa} in two important ways: (i) in considering
all five operators with dimension $\le 5$, and (ii) in pointing out the
effectiveness of the off-shell Higgs boson cross section for determining the
coupling structure.

We found that all non-SM operators lead to larger off-shell $gg \to X
\to ZZ^\ast \to 4\ell$ cross sections.  This will allow a
complementary constraint on (or measurement of) the 
non-SM couplings of the
putative Higgs boson to $Z$ bosons.  This is especially true for the
$\mathcal{O}_4$ operator; however, its amplitude violates unitarity at
relatively low energies (in a way that was quantified above).

Another way to interpret this result is to note that if experimental
tests of the invisible Higgs boson width in this channel, along the lines
suggested in Refs.~\cite{Caola:2013yja, Campbell:2013una,
  oai:arXiv.org:1206.4803, Kauer:2013cga, Campbell:2013wga,
  Passarino:2013bha}, observe an excess in high invariant mass
four-lepton events, then we will be presented with the challenge of 
determining whether this signal
results from the invisible width of the Higgs boson or from higher
dimensional operators, 
as both serve to enhance the off-shell Higgs boson cross section for a given
on-shell
Higgs boson cross section.  In fact, one could consider the parameter space
consisting of the coupling constants for the five operators,
$\kappa_{1-5}$, and the invisible width of the Higgs.  Limits on non-SM
$XZZ$ couplings from the
Higgs contribution to the off-shell four-lepton cross section are
strengthened by the addition of non-negligible invisible width for the Higgs.

We also noted that the ``contact operator'' $\mathcal{O}_5$ produces
very off-shell $Z$ bosons at large $\sqrt{\hat{s}}$.  While the cross
section for the production of these events is rather small at the LHC,
future colliders may be able to measure or constrain $\kappa_5$ using
this interesting effect.

Future work will include the effect of interference with the $gg
\to ZZ$ background explicitly, as this is the dominant effect in
constraining the magnitude of Higgs contributions to the four-lepton
cross section at large invariant mass.  It is particularly interesting
to see how the magnitude of this interference changes when
varying the $XZZ$ tensor structure.  Also of interest is the
effect on precision electroweak observables~\cite{Gori:2013mia, 
Chen:2013kfa} 
from the five operators considered above, as well as the natural
extension to other Higgs boson production processes such as weak
vector boson fusion or associated production.
We note that while the ``golden'' four-lepton channel has many
benefits, the framework provided here could be easily extended to
other channels, in particular, other channels which also involve $X\to
VV$ decays.  

\section{Acknowledgements}
We thank A.~Gritsan, A.~Korytov, I.~Low, F.~Maltoni, G.~Mitselmakher,
and C.~Williams for useful 
discussions.  
JG, JL, KM and SM thank their CMS colleagues.
JL acknowledges the hospitality of the SLAC Theoretical Physics Group.
MP is supported by the World Premier International Research Center
Initiative (WPI Initiative), MEXT, Japan. 
Work supported in part by U.S. Department of
Energy Grants DE-FG02-97ER41029.  Fermilab is operated by the Fermi
Research Alliance under contract DE-AC02-07CH11359 with the
U.S. Department of Energy.

\end{document}